\documentclass[twocolumn,prd,showpacs,nofootinbib]{revtex4}
\usepackage{amsmath,amsfonts,braket}
\begin{document}
\title{\bf Atomic and molecular transitions induced by axions via oscillating nuclear moments}
\author{V. V. Flambaum$^{1,2,3}$, H. B. Tran Tan$^{1}$, D. Budker$^{2,4}$ and A. Wickenbrock$^{2}$
} 
\affiliation{
$^1$
School of Physics, University of New South Wales,  Sydney 2052,  Australia}
\affiliation{$^2$Helmholtz Institute Mainz, Johannes Gutenberg-Universit\"at, 55099 Mainz, Germany}
\affiliation{$^3$The New Zealand Institute for Advanced Study, Massey University Auckland, 0632 Auckland, New Zealand}
\affiliation{$^4$Department of Physics, University of California, Berkeley, California 94720, USA}
\date{\today}

\begin{abstract}
The interaction of standard model's particles with the axionic Dark Matter field may generate oscillating nuclear electric dipole moments (EDMs), oscillating nuclear Schiff moments and oscillating nuclear magnetic quadrupole moments (MQMs) with a frequency corresponding to the axion's Compton frequency. Within an atom or a molecule an oscillating EDM, Schiff moment or MQM can drive transitions between atomic or molecular states. The excitation events can be detected, for example, via subsequent fluorescence or photoionization. Here we calculate the rates of such transitions. If the nucleus has octupole deformation or quadrupole deformation then the transition rate due to Schiff moment and MQM can be up to $10^{-16}$ transition per molecule per year. In addition, an MQM-induced transition may be of M2-type, which is useful for the elimination of background noise since M2-type transitions are suppressed for photons.
 \end{abstract}

\maketitle
\section{Introduction}\label{Introduction} 

The nature of dark matter (DM) remains unknown.
The axion is a prominent dark matter candidate originally introduced in the 1970s to explain the apparent charge-parity (CP) symmetry of the strong interactions. Most searches for axion and axion-like particles (ALPs)\footnote{ALPs are pseudoscalar particles like the axion that do not, however, solve the strong-CP problem; we refer to both axions and ALPs as `axions' in this paper.} rely on the conversion between axions and photons. Recently, experiments like the Cosmic Axion Spin Precession experiments (CASPEr) started to look for other types of axion couplings \cite{Casper}. Assuming that the dark matter in the Milky Way consists of axions , the dark matter field can be described as a field oscillating at a frequency close to the Compton frequency of the axion. This field induces oscillating electric dipole moments (EDMs), oscillating Schiff moments and an oscillating magnetic quadrupole moments (MQMs) of fundamental particles, nuclei, atoms, and molecules \cite{Stadnik2014,Graham}. These moments cause the precession of particle's spins due to gradients in the axion field (the axion-wind effect) \cite{GrahamRajendran} and may thus, in principle, be detected. The CASPEr experiments search for spin precession due to the nuclear Schiff moment, which is related to the nucleon EDMs and axion-induced P, T-odd nuclear forces, and axion wind with nuclear magnetic resonance. First results constraining the axion-nucleon couplings have been published by CASPEr \cite{Antoine, Comagnetometer}, as well as by other experiments re-analyzing existing data obtained in the neutron-EDM \cite{nEDM} and atomic co-magnetometer experiments \cite{VolanskyArxiv} and trapped anti-proton experiment\ \cite{smorra2019direct}.


In this paper, we analyze the effect of an axion-induced oscillating nuclear EDM, an oscillating nuclear Schiff moment and an oscillating nuclear MQM in atoms and molecules. Such oscillating moments may induce transitions in the atom or molecule if the transition frequency matches the axion oscillation frequency. In fact, nuclear-EDM-induced atomic and molecular transitions were already discussed in Ref.\ \cite{ArvanitakiMolecule}. In this work, we extend the discussion to the case of Schiff-moment-induced- and nuclear-MQM-induced transitions whose rates, as we will demonstrate, may be several orders of magnitude larger than that of an EDM-induced transitions. We present estimates of the corresponding transition rates, discuss their scaling with the relevant atomic and molecular parameters and comment on the feasibility of experimental observation of the effects.

\section{Nuclear EDM produced by the axion dark matter field}\label{Nuclear EDM} As noted in Ref.\ \cite{Witten}, a neutron EDM may be produced by the so-called `QCD $\theta$-term'. Numerous references and recent results for the neutron and proton EDMs are summarised in the review\ \cite{Yamanaka}
 \begin{equation}\label{theta}
 \begin{aligned}
 d_n&= -(2.7 \pm 1.2) \times 10^{-16} \theta \, e\,  \textrm{cm} \,,\\
 d_p&= (2.1 \pm 1.2) \times 10^{-16} \theta \, e\,  \textrm{cm}\,.
 \end{aligned}
\end{equation}

Calculations of the EDM  of a nucleus with more than nucleon produced by the P,T-odd nuclear forces were performed in Refs.\ \cite{HH,SFK,FKS1985,FKS1986,KhriplovichKorkin2000,Yoshinaga2010,MEREGHETTI2011,deVriesFormFactor2011,
deVriesEDM2011,WIRZBA2014,Dekens2014,YoshinagaXe2014,Bsaisou2015,DeVries,Yamanaka6Li,YamanakaLightNuclei,
YamanakaPhysRevC2018,Yamanaka:2019vec}. For a general estimate of such a nuclear EDM, it is convenient to use a single-valence-nucleon formula from Ref.\ \cite{SFK} and express the result in terms of $\theta$ following Ref.\ \cite{FDK}:
\begin{equation}\label{d}
   d \approx   7 \times 10^{-16}\left(q- \frac{Z}{A}\right)\left(1- 2q\right)\langle \sigma\rangle  \theta\, e\, {\rm cm}\,,
  \end{equation}
where $q=1$ for the valence proton, $q=0$ for the valence neutron, the nuclear spin matrix element is $\langle \sigma\rangle=1$ if $j=l+1/2$ and $\langle \sigma\rangle=-j/\left(j+1\right)$ if $j=l-1/2$. Here, $j$ and $l$ are the total and orbital momenta of the valence nucleon.

There are many specific results for the  $^2$H and $^3$He EDMs, see, e.g., reviews\ \cite{YamanakaLightNuclei,Chupp}. Within the error bars, the deuterium EDM is consistent with zero due to the cancellation between the proton and neutron contributions. The $^3$He nucleus contains an unpaired neutron. Using the calculation of the P,T-odd nuclear force contribution from Refs.\ \cite{Yamanaka6Li,Bsaisou2015,DeVries} ($-1.5 \times 10^{-16} \theta \, e\,  \textrm{cm}$) and the value of the neutron  EDM from Eq.\ \eqref{theta}, we obtain for the $^3$He EDM
\begin{equation}\label{3He}
   d(^3\textrm{He}) = (- 4.2 \pm 1.5)  \times 10^{-16}  \theta \, e  \, \textrm{cm}\,,
  \end{equation}  
which is close to the estimate obtained using Eq.\ \eqref{d}, which gives $d(^3\textrm{He}) = - 4.7  \times 10^{-16}  \theta \, e  \, \textrm{cm}$, even though the latter does not include the contribution of the neutron EDM.

The Schiff moment of a  nucleus is defined as\ \cite{SFK}
\begin{equation}\label{Schiff}
{\bf S}=\frac{e}{10}\left[\left<r^2{\bf r}\right>-\frac{5}{3Z}\left<r^2\right>\left<{\bf r}\right>\right]\,,
\end{equation}
where $\left<r^n\right>=\int\rho\left({\bf r}\right)r^nd^3r$ are the moments of the nuclear charge density and $\bf r$ is measured from the nuclear center of mass ($n=2,3$ is relevant in Eq.\ \eqref{Schiff}). For a spherical nucleus with one unpaired nucleon, Eq.\ \eqref{Schiff} becomes
\begin{equation}
S=-7\times10^{-7}\theta eq\left[\left(\langle\sigma\rangle+\frac{1}{I+1}\right)\langle r^2\rangle-\frac{5}{3}\langle\sigma\rangle r_q^2\right]\,,
\end{equation}
where $q$ and $\langle\sigma\rangle$ are defined as in Eq.\ \eqref{d} and $r^2_q$ is the mean square charge radius. Approximately, $\langle r^2\rangle\approx r_q^2\approx(3/5)R^2$ where $R=r_0A^{1/3}$ is the mean bound-state radius of the nucleus ($r_0\approx1.5\,{\rm fm}$ and A is the nuclear mass number).

It is observed from Eq.\ \eqref{Schiff} that the Schiff moment is a result of adding two terms of opposite sign. These two terms are often not known with sufficient accuracy so the result of calculating the Schiff moment becomes unreliable. Also, the Schiff moment is determined by the charge distribution of the protons. However, it is directed along the nuclear spin which, for example in ${}^{199}{\rm Hg}$, is carried by the valence neutron, so the Schiff moment is determined by the many-body effects which are harder to calculate. The calculation of nuclear EDMs, on the other hand, does not suffer from these problems (such as the difference of two nearly equal terms) and thus have certain computational advantages.

Despite of these difficulties, numerical calculations of the Schiff moments do exist. Some numerically computed values for the Schiff moments of spherical nuclei with one unpaired nucleon are\ \cite{SFK,FKS1985,FKS1986,FlambaumGinges2002,JesusHg2005}
\begin{equation}
\begin{aligned}
S\left({}^{205}{\rm Tl}\right)&\approx-7.4\times 10^{-3}\theta\, e\, {\rm fm}^3\,,\\
S\left({}^{199}{\rm Hg}\right)&\approx-4.8\times 10^{-3}\theta\, e\, {\rm fm}^3\,.
\end{aligned}
\end{equation}

We note in passing that there are experimental limits on the static Schiff moments of ${}^{205}{\rm Tl}$ and ${}^{199}{\rm Hg}$, e.g., $S_{{}^{199}{\rm Hg}}\lesssim 3.1 \times 10^{-13}e\, {\rm fm}^3$ corresponding to $\left|\theta\right|\lesssim 10^{-10}$\ \cite{ChoTl1991,GriffithHg2011,GranerHg2017}.

In nuclei with octupole deformation, the nuclear Schiff moments may be enhanced by 100 to 1000 times (thanks to  the small energy differences of nuclear levels with opposite parity and collective effects)\ \cite{Auberbach1996,Spevak1997,EngelRa2003,Dobaczewski2018,Flambaum2019tym,flambaum2019electric}
\begin{equation}
\begin{aligned}
S\left({}^{153}{\rm Eu}\right)&\approx 3.7\, \theta\, e\, {\rm fm}^3\,,\\
S\left({}^{225}{\rm Ra}\right)&\approx \theta\, e\, {\rm fm}^3\,,\\
S\left({}^{227}{\rm Ac}\right)&\approx 10\, \theta\, e\, {\rm fm}^3\,,\\
S\left({}^{229}{\rm Th}\right)&\approx 2\, \theta\, e\, {\rm fm}^3\,,\\
S\left({}^{235}{\rm U}\right)&\approx 3\, \theta\, e\, {\rm fm}^3\,,\\
S\left({}^{237}{\rm Np}\right)&\approx 6\, \theta\, e\, {\rm fm}^3\,.\\
\end{aligned}
\end{equation}

It should also be mentioned that for the Schiff moment of a nucleus with octuple deformation, the aforementioned problem with the cancellation of two nearly equal terms does not arise since the first term in Eq.\ \eqref{Schiff} is significantly enhanced whereas the second term is not (for similar distribution of protons and neutrons) and the result is stable. The calculation of the Schiff moment is, in this case, reduced to that of the expectation value of the P,T-odd interaction over intrinsic states of the deformed nucleus.

The MQM of a spherical nucleus with one unpaired nucleon is given by\ \cite{SFK,FDK,Skripnikov2017}
\begin{equation}
\hspace*{-0.15cm}M=\left[d-6\times 10^{-16}\theta\left(\mu -q\right)\left(e\,{\rm cm}\right)\right]\lambda_p\left(2I-1\right)\langle\sigma\rangle\,,
\end{equation}
where $d$ is the valence nucleon's EDM, $\mu$ is its magnetic dipole moment in nuclear magnetons, $q$, $\sigma$, $I$ are as defined in Eqs.\ \eqref{d} and\ \eqref{Schiff} and $\lambda_p=\hbar/m_pc=2.1\times 10^{-14}\,{\rm cm}$ ($m_p$ is the proton mass).

Some values for the MQM of spherical nuclei are\ \cite{SFK}
\begin{equation}
\begin{aligned}
M\left({}^{131}{\rm Xe}\right)&\approx -3\times 10^{-29}\theta\, e\, {\rm cm}^2\,,\\
M\left({}^{201}{\rm Hg}\right)&\approx 5\times 10^{-29}\theta\, e\, {\rm cm}^2\,.\\
\end{aligned}
\end{equation}

If the nucleus in question has quadrupole deformation, collective effects generally enhance the nuclear MQM by one order of magnitude\ \cite{FlambaumMQM1994,Skripnikov2017,FDK,Lackenby2018}
\begin{equation}
\begin{aligned}
M\left({}^{173}{\rm Yb}\right)&\approx -9\times 10^{-28}\theta\, e\, {\rm cm}^2\,,\\
M\left({}^{177}{\rm Hf}\right)&\approx -1\times 10^{-27}\theta\, e\, {\rm cm}^2\,,\\
M\left({}^{179}{\rm Hf}\right)&\approx -2\times 10^{-27}\theta\, e\, {\rm cm}^2\,,\\
M\left({}^{181}{\rm Ta}\right)&\approx -2\times 10^{-27}\theta\, e\, {\rm cm}^2\,,\\
M\left({}^{229}{\rm Hf}\right)&\approx -1\times 10^{-27}\theta\, e\, {\rm cm}^2\,.
\end{aligned}
\end{equation}

Reference\ \cite{Graham} discussed the possibility that the dark matter field is, in fact, an oscillating axion field which generates neutron EDMs. This axion field may also generate oscillating nuclear EDMs, oscillating nuclear Schiff moments and oscillating nuclear MQMs\ \cite{Stadnik2014}. Relating the value of the axion field to the local dark matter density (Ref.\ \cite{Stadnik2014}), we may substitute $\theta(t)=\theta _0 \cos(\omega t)$ where $\theta _0= 4 \times 10^{-18}$, $\omega=m_a c^2/\hbar$ and $m_a$ is the axion mass\footnote{It is worth mentioning that axions are a stochastic field with a finite coherence times $\tau_c\approx \frac{10^{-6}h}{m_ac^2}$, see, for example Ref.\ \cite{Centers:2019dyn}. In this paper, by $\theta _0$ we mean $\sqrt{\langle\theta _0^2\rangle}$.}. It is important to keep in mind that ALPs inducing larger dipole moments are also among viable DM candidates, so an experiment with sensitivity less than that necessary to detect axionic DM could already be sensitive to DM composed of such ALPs.


\section{Atomic transitions induced by oscillating nuclear moments}\label{Atom_transition}
We have presented in the last section the possibility of having oscillating nuclear EDMs, Schiff moments and MQMs. The interactions of these moments with the atomic electrons may cause electronic transitions. In what follows, we provide estimates of the transition rates due to the interactions with these moments.
\subsection{Nuclear EDM contribution}
The interaction of the atomic electrons with a nuclear EDM $\bf d$ may be presented as
\begin{equation}\label{Phi}
V_{\rm atom}^{\rm EDM}= 
e\sum_{k=1}^{N_e}\frac{{\bf d }\cdot {\bf r}_k}{r_k^3}=\frac{i}{Z e \hbar}[{\bf P \cdot d},H_0] \,,
\end{equation}
where $H_0$ is the (Schr\"odinger or Dirac) Hamiltonian for the atomic electrons in the absence of $\bf d$, $N_e$ is the number of the electrons, $Ze$ is the nuclear charge, $- e$ is the electron charge, ${\bf r}_k$ is  the electron position relative to the nucleus and ${\bf P}=\sum_{k=1}^{N_e}{\bf p}_{k}$ is the total momentum of all atomic electrons. 

The second equal sign in Eq.\ \eqref{Phi} holds because we have assumed that the nuclear mass is infinite and neglected the small effects of the Breit and magnetic interactions. The operator $\bf P$ commutes with the electron-electron interaction but does not commute with $U$, the potential energy due to the interaction with the nucleus
\begin{equation}
U=-\sum_{k=1}^{N_e}Ze^2/r_k\,,
\end{equation}
so we have
\begin{equation}\label{Commutator}
\begin{aligned}
 \hspace*{-0.11cm}\left[{\bf P },H_0\right] &= \left[{\bf P }, 
 U\right] 
 = -i \hbar  Z e^2 \sum_{k=1}^{N_e}\nabla \frac{1}{r_k}\\
 &=i\hbar Ze^2\sum_{k=1}^{N_e}\frac{{\bf r}_k}{r_k^3}\,.
\end{aligned}
\end{equation}

Using the nonrelativistic relation
\begin{equation}
{\bf P}=- \frac{i m} {e \hbar} [H_0,{\bf D}]\,,
\end{equation}
where $m_e$ is the electron's mass and ${\bf D}=-e \sum_{k=1}^{N_e} {\bf r}_k$ is the atomic electric dipole moment, the matrix element of $V_{\rm atom}^{\rm EDM}$ may be written as\ \cite{ArvanitakiMolecule}
\begin{equation}\label{M_EDM_atom}
\bra{f}V_{\rm atom}^{\rm EDM}\ket{i}=-\frac{\omega^2 m_e}{Ze^2} {\bf d}\cdot\bra{f}{\bf D }\ket{i}\,,
\end{equation}
where $\omega$ is the frequency of the axion field, which must matches the transition frequency $\left(E_f-E_i\right)/\hbar$.
 
The scalar operator $V_{\rm atom}^{\rm EDM}$ conserves the total atomic angular momentum $F$. For the electron variables the selection rules are identical to that for E1 amplitudes. 

In accordance with the Schiff theorem, which states that the static EDM of a subatomic point particle is unobservable in the nonrelativistic limit, the matrix element\ \eqref{M_EDM_atom} is propotional to $\omega^2$ and thus vanishes for $\omega=0$. This $\omega^2$ suppression should not appear for the transitions induced by oscillating Schiff moments and MQMs which we shall consider below. 

The transition probability $W \propto  \left|\bra{i}V_{\rm atom}^{\rm EDM}\ket{f}\right|^2 $ is inversely proportional to the squared nuclear charge,  $W \propto 1/ Z^2$, i.e., light atoms like H, He, Li are more advantageous for experiments. The origin of this factor may be related to the extended Schiff theorem for ion: the screening factor for an external electric field scales as $1/Z$.

Note that the transition probability is not suppressed for high electron waves $j,l$. The reason is that the  matrix element of $V \sim 1/r^2$ does not converge at small distances. Indeed, an estimate  of the contribution of the small distances $\int (\psi_1^+\psi_2 /r^2) r^2 dr = 
 \int \psi_1^+\psi_2  dr  \sim r^{l_1+l_2 +1}$ actually converges on the atomic size.    This is also the reason why the relativistic corrections are not important (except for in the values of energies $E_1,E_2$).  
 
Note also that the matrix element rapidly decreases with the electron principal quantum number $n$ since $\omega_{fi} \sim (n_f-n_i)/n^3$, $\bra{i}{\bf D}\ket{f} \sim n^2$, i.e., $\bra{i}V\ket{f} \sim 1/n^4$ (we assume that $n_f\approx n_i\approx n$). 
 
The probability of the transition on resonance for the stationary case (time $t \gg 1/\Gamma$ with $\Gamma$ being the dominant decoherence mechanism) for the perturbation $V_{\rm atom}^{\rm EDM}=V^0 \cos(\omega t) $ is given by the following expression\ \cite{Landau}:
\begin{equation}\label{rate}
W_{fi}=\frac{|\bra{f}V_{\rm atom}^{\rm EDM}\ket{i}|^2t}{\hbar \Gamma}  \,.
\end{equation}

In our case $\Gamma$ may be the width of the axion energy distribution ($\Gamma \sim 10^{-6} m_a c^2 =10^{-6}  \hbar \omega$) if the atomic level width is smaller than the axion energy distribution width (or the atomic energy level width $\Gamma$  in the opposite case). 
 
Inserting Eq.\ \eqref{M_EDM_atom} into Eq.\ \eqref{rate}, we obtain the approximate time for one transition to happen
\begin{equation}\label{t_EDM_atom} 
\begin{aligned}
t_{\rm atom}^{\rm EDM}&\approx \frac{2\times10^{22}}{N} Z^2\left(\frac{1 {\rm eV}}{\omega}\right)^3 \left|\frac{3 e a_B}{\bra{f}D_z\ket{i}}\right|^2\\
&\times\left(\frac{ 4 \times 10^{-16} \theta_0 \, e\,  \textrm{cm}}{d}\right)^2 \, \textrm{years}\,,
\end{aligned}
\end{equation}
where $N$ is the number of atoms and  $a_B$ is the Bohr radius. We have presented the result for the maximal projection of the atomic  angular momentum ($F_z=F=j+I$, where $j$ is the electron angular momentum and $I$ is the nuclear spin) and normalized the result to an atomic scale energy of 1 eV, a typical value of E1-amplitude $|\bra{f}D_z\ket{i}|=3 ea_B$ and a typical value of nuclear EDM $d_0=4 \times 10^{-16} \theta \, e\,  \textrm{cm}$. 

Note that for the hydrogen atom and transitions between highly excited Rydberg states of electron, there are analytical expressions \cite{Bethe} for the transition frequencies (between states with principal quantum numbers $n_i$ and $n_f$) $\omega \sim (n_i-n_f)/n^3$ and E1 amplitudes $\bra{f}D_z\ket{i} \sim n^2 a$. Altogether we have $t \propto Z^2 n^5$. 

\subsection{Nuclear Schiff moment contribution}\label{AtomSchiff}
The effective Hamiltonian for the interaction between a nuclear Schiff moment and an atomic electron is given by\ \cite{SFK}
\begin{equation}\label{Schiff_Hm}
V_{\rm atom}^{\rm SCHIFF}=4\pi{\bf S}\cdot\boldsymbol{\nabla}\rho\,,
\end{equation}
where ${\bf S}= S{\bf I}/I$ is the Schiff moment, $\bf I$ is the nuclear spin and $\rho$ is the nuclear density normalised to 1 (a better form for this interaction which takes into account the finite size of the nucleus is available, see Ref.\ \cite{FlambaumGinges2002}; for our estimate, the form\ \eqref{Schiff_Hm} is sufficient).

The matrix element for this interaction reads
\begin{equation}\label{M_Schiff_atom}
\begin{aligned}
\bra{f}V_{\rm atom}^{\rm SCHIFF}\ket{i}&=\langle f|4\pi\frac{\partial\rho}{\partial z}|i\rangle S\,,
\end{aligned}
\end{equation}
where we have assumed maximal projection of the nuclear spin and that the wavefunction of the system factorizes into nuclear and atomic parts. 

Inserting Eq.\ \eqref{M_Schiff_atom} into an analog of Eq.\ \eqref{rate}, we find the time for detection of one transition
\begin{equation}\label{t_Schiff_atom}
\begin{aligned}
t^{\rm SCHIFF}_{\rm atom}&\approx\frac{2\times 10^{22}}{N}\left(\frac{{\omega}}{1\rm eV}\right)\left(\frac{50000\frac{e}{a_B^{4}}}{\langle f|4\pi\frac{\partial\rho}{\partial z}|i\rangle}\right)^2\\
&\times\left(\frac{\theta_0\,e\,{\rm fm}^3}{S}\right)^2\, {\rm years}\,,
\end{aligned}
\end{equation}
where we have normalized the result to the atomic transition energy scale 1 eV, a typical value of the matrix element $\langle f|4\pi\frac{\partial\rho}{\partial z}|i\rangle \sim 50000a_B^{-4}$\ \cite{DzubaFlambaumGinges2000} and a typical value of the nuclear Schiff moment of a nucleus with octupole deformation $S \sim \theta_0\,e\,{\rm fm}^3$. If a maximal value of $S = 10\, \theta_0\,e\,{\rm fm}^3$ (${}^{227}{\rm Ac}$) was used then the result is two order of magnitudes better than\ \eqref{t_Schiff_atom}.

We observe that unlike $t^{\rm EDM}_{\rm atom}$, which is proportional to $\omega^{-3}$, $t^{\rm SCHIFF}_{\rm atom}\propto\omega$. This is because the matrix element\ \eqref{M_EDM_atom} is proportional to $\omega^2$ whereas the matrix element\ \eqref{M_Schiff_atom} does not depend on $\omega$. The factor $\omega$ in Eq.\ \eqref{t_Schiff_atom} comes from the width $\Gamma$ which is assumed to be $10^{-6}\omega$.

As noted in Sect.\ \ref{Nuclear EDM}, for a typical spherical nucleus with one unpaired nucleon, the Schiff moment is two to three orders of magnitude smaller than that in Eq.\ \eqref{t_Schiff_atom}. As a result, the `waiting' time if such nuclei are used is from four to six orders of magnitude larger than that in Eq.\ \eqref{t_Schiff_atom}.

Note also that the matrix element\ \eqref{M_Schiff_atom} is subjected to E1 selection rules (the matrix element $\bra{f}4\pi\frac{\partial \rho}{\partial z}\ket{i}$ contains the factor $\left( \begin{matrix}
   {{j}_{f}} & 1 & {{j}_{i}}  \\
   -{{m}_{f}} & 0 & {{m}_{i}}  \\
\end{matrix} \right)\xi \left( {{l}_{f}}+{{l}_{i}}+1 \right)$
where $\xi \left(x\right) =1 $ if $x$ is even and $\xi \left(x\right) =0 $ if $x$ is odd, see, e.g.,\ \cite{DzubaFlambaumGinges2000}).
\subsection{Nuclear magnetic quadrupole moment contribution}\label{NuclMQMAtm}
The interaction of an atomic electron with a nuclear MQM has the form\ \cite{SFK}
\begin{equation}
V_{\rm atom}^{\rm MQM} = -\frac{M}{4I\left(2I-1\right)}T_{ij}A_{ij}\,,
\end{equation}
(summation over the indices $i,j$ is implied) where the tensor $T_{ij}$ is defined by ($\bf I$ is again the nuclear spin)
\begin{equation}\label{Tensor T}
T_{ij}=I_iI_j+I_jI_i-\frac{2}{3}\delta_{ij}I\left(I+1\right)\,,
\end{equation}
and the tensor $A_{ij}$ by 
\begin{equation}
A_{ij}=\epsilon_{nmi}\alpha_n\partial_m\partial_j\frac{1}{r}\,.
\end{equation}
Here, $M$ is the magnitude of the MQM, $\alpha_n$ are the Dirac alpha matrices and $r$ the distance from the electron to the nucleus.

The matrix element of this interaction reads (assuming maximal projection of the nuclear spin)
\begin{equation}\label{M_MQM_atom}
\bra{f}V_{\rm atom}^{\rm MQM}\ket{i}=-\frac{I+1}{6\left(2I-1\right)}\bra{f}A_{zz}\ket{i}M\,,
\end{equation}
so the time for the detection of one transition is
\begin{equation}\label{t_MQM_atom}
\begin{aligned}
&t_{\rm atom}^{\rm MQM}\approx \frac{10^{22}}{N}\left(\frac{2I-1}{I+1}\right)^2\left(\frac{\omega}{1{\rm eV}}\right)\\
&\times\left(\frac{10\frac{e}{a^3_B}}{\langle f|A_{zz}|i\rangle}\right)^2\left(\frac{10^{-27}\theta_0\,e\,{\rm cm}^2}{M}\right)^2\, {\rm years}\,,
\end{aligned}
\end{equation}
where we have normalized the result to the atomic transition energy scale 1 eV, a typical value of the matrix element $\langle f|A_{zz}|i\rangle \sim 10\frac{e}{a_B^{3}} $\ \cite{DzubaFlambaumGinges2000} and a typical value of the nuclear MQM for a nucleus with quadrupole deformation $M \sim 10^{-27}\theta_0\,e\,{\rm cm}^2$. If the value $2\times10^{-27}\theta_0\,e\,{\rm cm}^2$ is used for $M$ (e.g., ${}^{179}$Hf) then the result is better by a factor of four. 

For a nucleus with no deformation, as noted in Sect.\ \ref{Nuclear EDM}, the MQM is generally an order of magnitude smaller than that presented in Eq.\ \eqref{t_MQM_atom}. As a result, the `waiting' time in such case is about two orders of magnitude larger than in Eq.\ \eqref{t_MQM_atom}.

Note that the matrix element\ \eqref{M_MQM_atom} is of M2 type (the matrix element $\bra{f}A_{zz}\ket{i}$ contains the factor $\left( \begin{matrix}
   {{j}_{f}} & 2 & {{j}_{i}}  \\
   -{{m}_{f}} & 0 & {{m}_{i}}  \\
\end{matrix} \right)\xi \left( {{l}_{f}}+{{l}_{i}}+1 \right)$). As a result, an MQM-induced transition, unlike an EDM-induced or a Schiff-moment-induced one, may be strongly suppressed for photons (in cases where E1-type transitions are impossible). This might prove useful for minimizing the effects of background processes.

\section{Molecular transitions induced by oscillating nuclear moments in diatomic molecule}\label{Mol_Trans}
In section\ \ref{Atom_transition}, we presented the estimates for the atomic transitions induced by various oscillating nuclear moments. These transitions are suitable for searching for axion with mass in the eV region. For axion mass in the $10^{-5}-10^{-1}$\,eV region (the value $m_a=26.2\, \mu {\rm eV}$ was recently predicted in Ref.\ \cite{AxionMass2017}; see also\ \cite{Buschmann:2019icd,Gorghetto_2018}), molecules appear to be better suited, with the region $10^{-5}-10^{-1}$\,eV coinciding with the typical separation between rovibrational states. In what follows, we present estimates for low-frequency transitions in diatomic molecules induced by oscillating nuclear moments. 
\subsection{Nuclear EDM contribution}

Let us now consider a diatomic molecule consisting of two nuclei of masses $M_I$, charges $Z_I$ and nuclear EDMs ${\bf d}_I$ (the subscript $I=1,2$
labels the nuclei) and $N_e$ electrons. As above, the position and momentum operators of the electrons in the laboratory frame are denoted by ${\bf r}_k$ and ${\bf p}_k$. The position and momentum operators of the
nuclei in the laboratory frame will be denoted by $\mathbf{R}_I$ and $\mathbf{P}_I$, respectively.

The interaction between the nuclear EDMs and the nuclei and electrons in the molecule may be presented as
\begin{equation}\label{Diatomic_Hamiltonian}
V_{\rm mol}^{\rm EDM}=\frac{\mathbf{d}_1\cdot \nabla_{{\bf R}_1}V_0}{Z_1e} + \frac{\mathbf{d}_2\cdot \nabla_{{\bf R}_2}V_0}{Z_2e}\,,
\end{equation}
where
\begin{equation}
V_0=\frac{Z_1Z_2e^2}{R_{12}} -\sum\limits_{k=1}^{N_e}\left(\frac{Z_1e^2}{R_{1k}}+\frac{Z_2e^2}{R_{2k}}\right)+\sum_{i\neq j}^{N_e}\frac{e^2}{r_{ij}}\,,
\end{equation}
is the Coulomb potential between the constituent particles of the molecule. Here, $R_{Ik}=\left| \mathbf{R}_I-\mathbf{r}_k \right|$,
$r_{hk}=\left| \mathbf{r}_h-\mathbf{r}_{k} \right|$ and
$R_{IJ}=\left| \mathbf{R}_{I}-\mathbf{R}_J \right|$. 

As shown in Appendix A, the matrix element of $V_{\rm mol}^{\rm EDM}$ has the form
\begin{equation}\label{EDM matrix element}
\bra{f}V_{\rm mol}^{\rm EDM}\ket{i} =\frac{{{\omega }^{2}}{{\mu }_{N}}}{e\sqrt{Z_1Z_2}} \bra{f}\boldsymbol{\delta }\cdot \mathbf{X} -\boldsymbol{\Delta }\cdot\sum\limits_{i=1}^{{{N}_{e}}}{{{\mathbf{x}}_{i}}}\ket{i}\,,
\end{equation}
where $\mathbf{X}=\mathbf{R}_1-\mathbf{R}_2$ is the inter-nuclear distance, ${{\mathbf{x}}_k}={{\mathbf{r}}_k}-\left(M_1{{\mathbf{R}}_{1}}+M_2{{\mathbf{R}}_{2}}\right)/\left(M_1+M_2\right)$ is the relative position of the electrons with respect to the nuclear center of mass and $\mu_N=M_1M_2/\left(M_1+M_2\right)$ is the reduced nuclear mass. The moments $\boldsymbol{\delta}$ and $\bf \Delta$ are defined as
\begin{equation}
\boldsymbol{\delta }=\sqrt{Z_1Z_2}\left(\frac{{\bf d}_1}{Z_1}-\frac{{\bf d}_2}{Z_2}\right)\,,
\end{equation}
and
\begin{equation}\label{Delta}
\begin{aligned}
\mathbf{\Delta }\approx \frac{{{m}_{e}}\left( {{M}_{1}}{{Z}_{2}}{{\mathbf{d}}_{1}}+{{M}_{2}}{{Z}_{1}}{{\mathbf{d}}_{2}} \right)}{{{M}_{1}}{{M}_{2}}\sqrt{Z_1Z_2}}\,,
\end{aligned}
\end{equation}
where in Eq.\ \eqref{Delta} we retain only the lowest-order term in the small parameters $m_e \ll M_{1,2}$. Here, $\omega$ is again the frequency of the axion field which matches the transition frequency $\left(E_f-E_i\right)/\hbar$.

To estimate the matrix element\ \eqref{EDM matrix element}, it is convenient to use the Born-Oppenheimer approximation, in which the molecular wavefunction can be written as
\begin{equation}\label{BOwavefunction}
\begin{aligned}
{{\psi }_{n}}&=\sqrt{\frac{2{{J}_{n}}+1}{8{{\pi }^{2}}}}D_{{{m}_{n}}{{\Omega }_{n}}}^{{{J}_{n}}}\left( \Theta  \right)\\
&\times\frac{\phi _{n}^{\text{vib}}\left( X \right)\varphi _{{{\kappa }_{n}}{{\Omega }_{n}}}^{e}\left( X,{{\mathbf{s}}_{i}} \right)}{X}\,,
\end{aligned}
\end{equation}
where $\sqrt{\left(2J+1\right)/\left(8\pi^2\right)}D^J_{m\Omega}$, $\phi^{\rm vib}/X$ and $\varphi^e_{\mu\Lambda}$ are, respectively, the rotational, vibrational, electronic wavefunctions. Here $J$ is the molecule's total angular momentum, $m$ is the projection of $J$ on a quantization axis (it is customary to use $M$ for this quantity; here we use $m$ to avoid confusion with the nuclear MQM), $\Omega$ is the projection of the electrons' total angular momentum on the molecular axis $\bf X$ and $\kappa$ stands for any other quantum numbers not listed so far. The rotational wavefunction $D^J_{m\Omega}\left(\Theta\right)$ is the Wigner $D$ matrix depending on the set of Euler angles $\Theta$ which descibes the molecule's orientation with respect to some laboratory-fixed frame.

As mentioned, an advantage of molecules over atoms is the existence of rovibrational states whose small energy spacing is convenient for searching for sub-eV dark matter. In this paper, we provide estimates for the transitions between such states. Also, in anticipation of the parity selection rules for the EDM, Schiff and MQM operators (see Sect.\ \ref{Atom_transition}), we assume that the states $f$ and $i$ have opposite parities, that is, $\ket{f}=\left(\ket{J',m,\Omega}-\left(-1\right)^{J'-\Omega}\ket{J',m,-\Omega}\right)/\sqrt{2}$ and $\ket{i}=\left(\ket{J,m,\Omega}+\left(-1\right)^{J-\Omega}\ket{J,m,-\Omega}\right)/\sqrt{2}$ (here we assume that $\Omega\neq 0$, the result for the case where $\Omega=0$ is similar). We also assume that $f$ and $i$ have the same electronic configuration.

Note that due to the small parameters $m_e/M_{1,2}$, we can assume for the moment that $|\boldsymbol{\Delta}|\ll |\boldsymbol{\delta}|$. Thus, if $\bra{f}{\bf X}\ket{i}\neq0$ then the term $\boldsymbol{\Delta }\cdot\sum\limits_{i=1}^{{{N}_{e}}}{{{\mathbf{x}}_{i}}}$ may be dropped from Eq.\ \eqref{EDM matrix element}. Also, if we are only interested in transitions between states within a single electronic configuration then the term $\boldsymbol{\Delta }\cdot\sum\limits_{i=1}^{{{N}_{e}}}{{{\mathbf{x}}_{i}}}$ is zero. In such cases, we are left with
\begin{equation}\label{M_EDM_mol}
\begin{aligned}
\bra{f}V_{\rm mol}^{\rm EDM}\ket{i} &\approx\frac{{{\omega }^{2}}{{\mu }_{N}}}{{{e}}} \bra{f}\boldsymbol{\delta }\cdot \mathbf{X}\ket{i}\\
  &={{\left( -1 \right)}^{2J'-m-\Omega}}B^{J'm\Omega}_{Jm\Omega}\frac{\omega^2\mu_N\delta{X}_{fi}}{e\sqrt{Z_1Z_2}}\,, \\ 
\end{aligned}
\end{equation}
(assuming maximal projection for $\boldsymbol{\delta}$) where
\begin{equation}\label{CJMO}
\begin{aligned}
B^{J'm\Omega}_{Jm\Omega}&=\sqrt{\left( 2J'+1 \right)\left( 2J+1 \right)}\\
&\times{{\left( \begin{matrix}
   J' & 1 & J  \\
   -m & 0 & m  \\
\end{matrix} \right)}}{{\left( \begin{matrix}
   J' & 1 & J  \\
   -\Omega  & 0 & \Omega   \\
\end{matrix} \right)}}\,,
\end{aligned}
\end{equation}
and
\begin{equation}
X_{fi}=\int{\bar{\phi }_{f}^{\text{vib}}X\phi _{i}^{\text{vib}}dX}\,.
\end{equation}
We note that in the case where ${\bf d}_2=0$, the result\ \eqref{M_EDM_mol} reduces to that presented in Ref.\ \cite{ArvanitakiMolecule}.

The time for the detection of one transition due to an oscillating nuclear EDM is (assuming that the dominant contribution to the width is due to the axion frequency distribution $\Gamma\approx10^{-6}\hbar\omega$) is
\begin{equation}\label{t_EDM_mol}
\begin{aligned}
t_{\text{mol}}^{\text{EDM}}&\approx \frac{6\times{{10}^{30}}}{N}\frac{{{Z}_{1}}{{Z}_{2}}}{\left(B^{J'm\Omega}_{Jm\Omega}\right)^2}{{\left( \frac{{{m}_{p}}}{{{\mu }_{N}}} \right)}^{2}}\left(\frac{\omega_\Omega}{\omega}\right)^3\\
&\times{{\left( \frac{3{{a}_{B}}}{{X_{fi}}} \right)}^{2}}{{\left( \frac{4\times {{10}^{-16}}\theta_0 \text{ }e\times \text{cm}}{\delta } \right)}^{2}}\,{\rm year}\,,
\end{aligned}
\end{equation}
where we have normalized our result to a typical $\Omega$-doublet separation $\omega_\Omega=10^{-5}{\rm eV}$, a typical inter-nuclear distance of $3a_B$ and a typical value for nulear EDM $4\times {{10}^{-16}}\theta_0 \text{ }e\times \text{cm}$.

Comparing Eqs.\ \eqref{M_EDM_atom} and\ \eqref{M_EDM_mol}, we see that for the same transition energy in the region of $10^{-5}\,{\rm eV}$, the molecular transition probability is enhanced by a factor of $\left(\mu_N/m_e\right)^2 \gtrsim 10^8$. This enhancement factor arises from the fact that because the nuclei mover slower than the electrons, the field created by the oscillating EDM of one nucleus on the other one is only partially screened by the electrons, see Ref.\ \cite{TranFlambaum2019}. Note also that the matrix element\ \eqref{M_EDM_mol} is subjected to E1 selection rules.

\subsection{Nuclear Schiff moment contribution}
The effective Hamiltonian for the interaction between the nuclear Schiff moment and the molecular axis ${\bf N}={\bf X}/X$ of a diatomic molecule is given by\ \cite{SFK}
\begin{equation}
V_{\rm mol}^{\rm SCHIFF} = W_S{\bf S}\cdot{\bf N}\,,
\end{equation}
where $W_S$ is the interaction constant.

Using the Born-Oppenheimer wavefunction\ \eqref{BOwavefunction} and assuming maximal projection of the nuclear spin, we obtain
\begin{equation}
\begin{aligned}
\bra{f}V_{\rm mol}^{\rm SCHIFF}\ket{i}\approx{{\left( -1 \right)}^{2{{J'}}-{{m}}-{{\Omega }}}}B^{J'm\Omega}_{Jm\Omega}W_SS\,, \\ 
\end{aligned}
\end{equation}
where $B^{J'm\Omega}_{Jm\Omega}$ is defined as in Eq.\ \eqref{CJMO}. The time for the detection of one transition due to an oscillating nuclear Schiff moment is thus
\begin{equation}\label{t_Schiff_mol}
\begin{aligned}
t^{\rm SCHIFF}_{\rm mol}&\approx\frac{2\times10^{17}}{N}\frac{1}{\left(B^{J'm\Omega}_{Jm\Omega}\right)^2}\frac{\omega}{\omega_\Omega}\\
&\times\left(\frac{50000\,\frac{e}{a_B^4}}{W_S}\right)^2\left(\frac{\theta_0\, e\,{\rm fm}^3}{S}\right)^2\,{\rm years}\,,
\end{aligned}
\end{equation}
where $W_S$ is normalized to a typical value of $50000\frac{e}{a_B^4}$\ \cite{Flambaum229Th2019} and $S$ to the typical value of $\theta_0\, e\, {\rm fm}^3$ (for a nucleus with octupole deformation). Using the maximal value of $S = 10\, \theta_0\,e\,{\rm fm}^3$ (${}^{227}{\rm Ac}$) results in a two orders of magnitudes shorter time than that of Eq.\,\eqref{t_Schiff_mol}.

We observe that, for the benchmark values of the parameters shown in Eqs.\ \eqref{t_EDM_mol} and\ \eqref{t_Schiff_mol}, the result\ \eqref{t_Schiff_mol} is thirteen orders of magnitude better than the result\ \eqref{t_EDM_mol}. If nuclei with no deformation are used instead, the results are still seven to nine order of magnitude better than those of using EDMs. Hence, for small axion frequency ($\omega\lesssim 10^{-3}{\rm eV}$), Schiff moments appear to be more advantageous than EDMs.

\subsection{Nuclear magnetic quadrupole moment contribution}\label{NuclMQMMol}
The effective Hamiltonian for the interaction between the nuclear MQM and the molecule is given by\ \cite{SFK}
\begin{equation}
V_{\rm mol}^{\rm MQM} = -\frac{W_MM}{2I\left(2I-1\right)}{\bf S}'\hat{\bf T}{\bf N}\,,
\end{equation}
where ${\bf S}'=\left(S'_{\xi},S'_{\psi},S'_{\zeta}\right)$ is an effective spin defined by the equations $S'_{\zeta}\ket{\Omega}=\Omega\ket{\Omega}$, $S'_{\pm}\ket{\pm\Omega}=0$, $S'_{\pm}\ket{\mp\Omega}=\ket{\pm\Omega}$ and $S'_{\pm}=S'_{\xi}\pm S'_{\psi}$ where the coordinates $\left(\xi,\psi,\zeta\right)$ form the molecular frame of reference with the $\zeta$-axis directed along the vector ${\bf N}$\ \cite{Kozlov_1987,DMITRIEV1992,Kozlov_1995}. The tensor $\hat{\bf T}$ is defined as in Eq.\ \eqref{Tensor T}. The factor $W_M$ is the coupling constant of the MQM interaction.

The matrix element of this interaction (assuming maximal projection of the nuclear spin) is
\begin{equation}\label{MQM matrix element}
\begin{aligned}
\bra{f}V_{\rm mol}^{\rm MQM}\ket{i}&\approx\left(-1\right)^{2J'-m-\Omega}C^{{J}'{m}{\Omega }}_{Jm\Omega}W_MM\,,
\end{aligned}
\end{equation}
where
\begin{equation}\label{DJMW}
\begin{aligned}
  C_{Jm\Omega }^{{J}'{m}{\Omega }}&=-\left(2/3\right) \sqrt{\left( 2{J}'+1 \right)\left( 2J+1 \right)} \\ 
 & \times \left( \begin{matrix}
   {{J}'} & 2 & J  \\
   -{m} & 0 & m  \\
\end{matrix} \right){\left( \begin{matrix}
   {{J}'} & 2 & J  \\
   -{\Omega } & 0 & \Omega   \\
\end{matrix} \right)}\,. \\ 
\end{aligned}
\end{equation}

Thus, the time for the detection of one transition due to an oscillating nuclear magnetic quadrupole moment is
\begin{equation}\label{t_MQM_mol}
\begin{aligned}
t_{\rm mol}^{\rm MQM} &\approx \frac{8\times10^{17}}{N}\frac{1}{\left(C^{J'm\Omega}_{Jm\Omega}\right)^2}\left(\frac{\omega}{\omega_\Omega}\right) \\
&\times\left(\frac{10^{33}\,\frac{\rm Hz}{e\,{\rm cm}^2}}{W_M}\right)^2\left(\frac{10^{-27}\theta_0\,e\,{\rm cm}^2}{M}\right)^2\,{\rm years}\,,
\end{aligned}
\end{equation}
where $W_M$ is normalized to a typical value of $10^{33}\frac{\rm Hz}{e{\rm cm}^2}$\ \cite{FDK} and $M$ to the typical value of $10^{-27}\theta_0\,e\,{\rm cm}^2$ (for a nucleus with quadrupole deformation). If the value $2\times10^{-27}\theta_0\,e\,{\rm cm}^2$ is used for $M$ (e.g., ${}^{179}$Hf) then the result is better by a factor of four. 

We observe that, for the benchmark values of the parameters, the result\ \eqref{t_MQM_mol} is thirteen orders of magnitude better than\ \eqref{t_EDM_mol}. If the nucleus in question has no quadrupole deformation, the result is still about eleven orders of magnitude better than using EDMs. Note also that the MQM-induced transition\ \eqref{MQM matrix element} is of M2 type, which is convenient for suppressing photon backgrounds (if $J'=J\pm 2$).

\section{Discussion and Conclusion}

In this paper, we presented the possibility of searching for axionic dark matter by means of atomic and molecular transitions induced by oscillating nuclear EDMs, nuclear Schiff moments and nuclear MQMs. While the search would be limited to axion frequencies closely corresponding to resonant atomic or molecular transitions, the latter can be tuned by using Zeeman and Stark effects. In addition, the transitions are `dense' in the region of Rydberg excitations, so complete coverage of a frequency interval is, in principle, possible. Molecules also have dense spectra and may present additional advantages for this kind of
experiments. 

From our estimates, it appears that for a transition frequency $\omega\sim 1\, {\rm eV}$, the contribution due to the nuclear Schiff moment (in a nucleus with octupole deformation) and nuclear MQM are comparable to that of the EDM.

On the other hand, in molecular rovibrational transitions, the contribution due to Schiff moment (in a nucleus with octupole deformation) and MQM is many orders of magnitude larger than that due to EDM. This is because MQM and Schiff moment matrix elements are not suppressed by the factor $\omega^2$ like EDM matrix elements (in this respect, atoms with small-$\omega$ transitions suffer from the same suppression of the EDM matrix elements). Also, an MQM-induced transition, unlike an EDM-induced or a Schiff-moment-induced one, which must be of E1 type and hence susceptible to photon background, can be of M2 type. Since M2-type transitions are strongly suppressed for photons, the effects of background processes may be limited. Thus, overall, MQMs are better suited for the detection of axions using atomic and molecular transition.

We note that the smallness of the rotational transition frequencies in molecules may pose a problem for the detection of axion signals if one is interested in this frequency range, with black body radiation potentially dominating the counts in the photon detector used to capture fluorescence light emitted when the molecules excited by axion-induced nuclear moments decay. This issue might be overcome by further exciting the excited molecules to a higher electronic state and then detecting the fluorescence emitted when the molecules in this final state relax back to the ground state. Since the frequency of an electronic transition is typically in the eV range, the fluorescence signal registered in the photon detector will not be dominated by black body radiation counts. We note also that this problem may arise only in the detection stage of the experiment. If the initial transition is induced by a nuclear MQM and is of $M2$-type then, as mentioned earlier, the background effects at this stage may be effectively suppressed. 

Finally, we point out that in our considerations, we implicitly assumed that the atoms and molecules in the sample are originally in the ground state. If the sample is in the gaseous phase at a temperature $T\sim 100\,{\rm K}$ (we focus on gases to avoid complications arising from the modification of atomic and molecular spectra in liquids and in solids; molecules in noble gas matrices are also a possible candidate) this assumption is reasonable if one targets the axion mass and thus axion-induced transition frequency $\omega\gtrsim 0.1\, {\rm eV}$, i.e., using vibrational or electronic transitions (thermal population of the states which lie at $\gtrsim 0.1\, {\rm eV}$ is small at this temperature). If, however, one is interested in scanning the small axion mass regime $\omega\sim 10^{-6}-10^{-3}\,{\rm eV}$ with molecules then at $T\sim 100\,{\rm K}$, most of the low-lying rotational states have non-negligible thermal populations, which poses a challenge for axion signal discrimination. One approach to improve upon this situation is to use trapped molecules which may be cooled down to $\mu{\rm K}$ or even $nK$, assuring that only the ground state is populated. The drawback of this approach is that the relatively small number of trapped molecules $N$ may degrade the statistical sensitivity to the axion. More modern technologies may some day allow us to cool a sufficiently large number particle down to a sufficiently small internal temperatures (see, e.g.,\ \cite{DimaCooling} and the references therein). Still the central experimental issue will be separating the axion related signal from the backgrounds, Here, two powerful tools are available for discriminating the useful signal: the M2 character of the axion/MQM induced transition as discussed above and the fact that the axion related signal should represent a sharp resonance with a relative frequency spread of a part per million or smaller, while the photon backgrounds are expected to be broad and non-resonant. 

We note finally that the relatively long coherence time of the axion suggests another possibility, in principle, to discriminate axion signals. Here one would apply a resonant laser field weakly driving the transition that is supposed to be driven by the axion-induced oscillating nuclear moments. The probability of the pure photon-induced transition does not depend on the phase of the driving field. However, the axion- and photon-induced amplitudes interfere, and the interference term should reverse with an appropriate phase reversal of the photon field. Since the axion phase is `reset' every coherence time, the overall interference signature should correspondingly change on the coherence-timescale of the axion. This peculiar signature is generally not expected from the background signal. An additional benefit of the interference scheme is an enhancement of the signal: the interference term in the transition probability is bilinear in the axion- and photon-induced amplitudes and the latter can be made reasonably large. A brief discussion of interference is presented in Appendix B.


\section*{Acknowledgment}
    The authors thank Mikhail G. Kozlov, Derek F. Jackson Kimball, Alexander O. Sushkov and Igor. B. Samsonov for helpful discussions. This work is supported by the Australian Research Council, the Gutenberg Fellowship and the New Zealand Institute for Advanced Study. It has also received support from the European Research Council (ERC) under the European Union Horizon 2020 Research and Innovation Program (grant agreement No. 695405), from the DFG Reinhart Koselleck Project and the Heising-Simons Foundation.
    
\section*{Appendix A}

In this appendix, we provide the derivation for Eq.\ \eqref{EDM matrix element}.

The total Hamiltonian of a diatomic molecule with the interaction with nuclear EDMs included is given by
\begin{equation}
\begin{aligned}
  H&=\frac{\mathbf{P}_{1}^{2}}{2{{M}_{1}}}+\frac{\mathbf{P}_{2}^{2}}{2{{M}_{2}}}+\sum\limits_{i=1}^{{{N}_{e}}}{\frac{\mathbf{p}_{i}^{2}}{2{{m}_{e}}}}\\
  &+{{V}_{0}}+{{V}^{\text{EDM}}_{\rm atom}} \,,\\ 
  {{V}_{0}}&=-\sum\limits_{i=1}^{{{N}_{e}}}{\frac{{{Z}_{1}}{{e}^{2}}}{\left| {{\mathbf{r}}_{i}}-{{\mathbf{R}}_{2}} \right|}}-\sum\limits_{i=1}^{{{N}_{e}}}{\frac{{{Z}_{2}}{{e}^{2}}}{\left| {{\mathbf{r}}_{i}}-{{\mathbf{R}}_{2}} \right|}}\\
  &+\sum\limits_{i>j}^{{{N}_{e}}}{\frac{{{e}^{2}}}{\left| {{\mathbf{r}}_{i}}-{{\mathbf{r}}_{j}} \right|}}+\frac{{{Z}_{2}}{{Z}_{1}}{{e}^{2}}}{\left| {{\mathbf{R}}_{2}}-{{\mathbf{R}}_{1}} \right|} \,,\\ 
  {{V}^{\text{EDM}}_{\rm atom}}&=\frac{{{\mathbf{d}}_{1}}\cdot {{\nabla }_{{{\mathbf{R}}_{1}}}}{{V}_{0}}}{{{Z}_{1}}e}+\frac{{{\mathbf{d}}_{2}}\cdot {{\nabla }_{{{\mathbf{R}}_{2}}}}{{V}_{0}}}{{{Z}_{2}}e}\,, \\ 
\end{aligned}
\end{equation}
where the nuclear positions ${\bf R}_{1,2}$, nuclear momenta ${\bf P}_{1,2}$, electrons position ${\bf r}_i$ and electron momenta ${\bf p}_i$ are defined in the laboratory frame.

A change of coordinates to the center-of-mass frame as described in Ref.\ \cite{TranFlambaum2019}, gives, after discarding the free motion of the molecule
\begin{equation}
\begin{aligned}
   H&={{H}_{0}}+{{V}^{\text{EDM}}_{\rm atom}} \,,\\ 
  {{H}_{0}}&=\frac{{{\mathbf{Q}}^{2}}}{2{{\mu }_{N}}}+\sum\limits_{i=1}^{{{N}_{e}}}{\frac{\mathbf{q}_{i}^{2}}{2{{\mu }_{e}}}}+\sum\limits_{i\ne j}^{{{N}_{e}}}{\frac{{{\mathbf{q}}_{i}}{{\mathbf{q}}_{j}}}{{{M}_{N}}}+{{V}_{0}}} \,,\\ 
\end{aligned}
\end{equation}
where $V_0$ and $V^{\text{EDM}}_{\rm atom}$ are now functions of the new variables ${\bf X}= {\bf R}_1-{\bf R}_2$ and ${\bf x}_i={\bf r}_i-\left(M_1{\bf R}_1+M_2{\bf R}_2\right)/M_N$. The momenta $\bf Q$ and ${\bf q}_i$ are conjugate to ${\bf X}$ and ${\bf x}_i$, respectively. For convenience, we have defined $M_N=M_1+M_2$, $M_T=M_N+N_em_e$, $\mu_N=M_1 M_2/M_N$ and $\mu_e=M_N m_e/\left(M_N+m_e\right)$.

We may then write
\begin{equation}
\begin{aligned}\label{V_EDM}
   {{V}^{\text{EDM}}_{\rm atom}}&=\frac{{{\mathbf{d}}_{1}}\cdot {{\nabla }_{{{\mathbf{R}}_{1}}}}{{V}_{0}}}{{{Z}_{1}}e}+\frac{{{\mathbf{d}}_{2}}\cdot {{\nabla }_{{{\mathbf{R}}_{2}}}}{{V}_{0}}}{{{Z}_{2}}e}\\
   &=\frac{i{{\mathbf{d}}_{1}}\cdot \left[ {{\mathbf{P}}_{1}},{{H}_{0}} \right]}{{{Z}_{1}}e\hbar }+\frac{i{{\mathbf{d}}_{2}}\cdot \left[ {{\mathbf{P}}_{2}},{{H}_{0}} \right]}{{{Z}_{2}}e\hbar } \\ 
 & =\frac{i{{\mathbf{d}}_{1}}\cdot \left[ \mathbf{Q}-\frac{{{M}_{1}}}{{{M}_{N}}}\sum\limits_{i=1}^{{{N}_{e}}}{{{\mathbf{q}}_{i}}},{{H}_{0}} \right]}{{{Z}_{1}}e\hbar }\\
 &-\frac{i{{\mathbf{d}}_{2}}\cdot \left[ \mathbf{Q}+\frac{{{M}_{2}}}{{{M}_{N}}}\sum\limits_{i=1}^{{{N}_{e}}}{{{\mathbf{q}}_{i}}},{{H}_{0}} \right]}{{{Z}_{2}}e\hbar }\,, \\ 
\end{aligned}
\end{equation}
so
\begin{equation}
\begin{aligned}
\bra{f}V^{\text{EDM}}_{\rm atom}\ket{i}&=-\frac{i\omega{{\mathbf{d}}_{1}}\cdot \bra{f} \mathbf{Q}-\frac{{{M}_{1}}}{{{M}_{N}}}\sum\limits_{i=1}^{{{N}_{e}}}{{{\mathbf{q}}_{i}}}\ket{i}}{{{Z}_{1}}e\hbar }\\
 &+\frac{i\omega{{\mathbf{d}}_{2}}\cdot \bra{f} \mathbf{Q}+\frac{{{M}_{2}}}{{{M}_{N}}}\sum\limits_{i=1}^{{{N}_{e}}}{{{\mathbf{q}}_{i}}}\ket{i}}{{{Z}_{2}}e\hbar }\,. \\ 
\end{aligned}
\end{equation}

Using the relations
\begin{equation}
\begin{aligned}
\mathbf{Q}&+\left( -{{1}^{I}} \right)\frac{{{M}_{I}}}{{{M}_{N}}}\sum\limits_{i=1}^{{{N}_{e}}}{{{\mathbf{q}}_{i}}}\\
&=\frac{i}{\hbar }\left[ {{H}_{0}},{{\mu }_{N}}\mathbf{X}+\left( -{{1}^{I}} \right)\frac{{{M}_{I}}{{\mu }_{e}}}{{{M}_{T}}}\sum\limits_{i=1}^{{{N}_{e}}}{{{\mathbf{x}}_{i}}} \right]
\end{aligned}
\end{equation}
we obtain
\begin{equation}
\bra{f}V_{\rm mol}^{\rm EDM}\ket{i} =\frac{{{\omega }^{2}}{{\mu }_{N}}}{e\sqrt{Z_1Z_2}} \bra{f}\boldsymbol{\delta }\cdot \mathbf{X} -\boldsymbol{\Delta }\cdot\sum\limits_{i=1}^{{{N}_{e}}}{{{\mathbf{x}}_{i}}}\ket{i}\,,
\end{equation}
where
\begin{equation}
\boldsymbol{\delta }=\frac{{{Z}_{2}}{{\mathbf{d}}_{1}}-{{Z}_{1}}{{\mathbf{d}}_{2}}}{\sqrt{{{Z}_{1}}{{Z}_{2}}}}\,,
\end{equation}
and
\begin{equation}
\begin{aligned}
\mathbf{\Delta }&=\frac{{{\mu }_{e}}\left( {{M}_{1}}{{Z}_{2}}{{\mathbf{d}}_{1}}+{{M}_{2}}{{Z}_{1}}{{\mathbf{d}}_{2}} \right)}{{{M}_{T}}{{\mu }_{N}}\sqrt{{Z}_{1}{Z}_{2}}}\\
&\approx \frac{{{m}_{e}}\left( {{M}_{1}}{{Z}_{2}}{{\mathbf{d}}_{1}}+{{M}_{2}}{{Z}_{1}}{{\mathbf{d}}_{2}} \right)}{{{M}_{1}}{{M}_{2}}\sqrt{Z_1Z_2}}\,.
\end{aligned}
\end{equation}

\section{Appendix B}

In this appendix, we briefly discuss the possibility of using interference to assist the detection of rotational transitions driven by axion-induced oscillating nuclear MQM.

The photon-induced MQM transition amplitude is given by
\begin{equation}\label{photonM2}
M_{\gamma}=\frac{\sqrt{\pi n_{\gamma}}Ze\omega^{3/2}}{\sqrt{2}Am_p}\bra{f}\left(\hat{\bf k}\cdot{\bf r}\right)\hat{\bf b}\cdot\left(g_R{\bf J}+g_N{\bf I}\right)\ket{i}\,,
\end{equation}
where $n_{\gamma}$ is the photon number density, $\hat{\bf k}$ is the photon propagation unit vector, $\hat{\bf b}=\hat{\bf k}\times\hat{\boldsymbol{\epsilon}}$ is the direction of the magnetic component of the photon field, $\bf J$ is the nucleus' total angular momentum, $\bf I$ is the nuclear spin and $g_R$ and $g_N$ are the $g$-factor associated with $\bf J$ and $\bf I$. We estimate the transition amplitude\ \eqref{photonM2} as
\begin{equation}\label{gamma_matrix}
M_{\gamma}\sim\frac{\sqrt{n_{\gamma}}Ze\omega^{3/2}\bar{X}}{Am_p}\,.
\end{equation}
Note that the photon density may be expressed in terms of the number of photons emitted by the source per unit time $P$ and the cross section $\sigma_{\gamma}$ of the photon beam as $n=P/\left(c\sigma_{\gamma}\right)$.

The ratio between the change in the signal as the axion phase resets and the total signal is
\begin{equation}\label{signal}
r\approx4\left|{M_a}/{M_{\gamma}}\right|\,,
\end{equation}
where $M_a$ is the MQM amplitude\ \eqref{MQM matrix element}. The factor of 4 appears because the cross term contains a factor of 2 and the effect of reversing the axion phase contains another factor of 2.

Substituting Eqs.\ \eqref{MQM matrix element} and\ \eqref{gamma_matrix} into Eq.\ \eqref{signal}, we obtain
\begin{equation}
\begin{aligned}
r&\approx10^{-4}\frac{A}{Z}\left(\frac{10^{-4}\,{\rm eV}}{\omega}\right)^{3/2}\left(\frac{10^{20}/{\rm s}}{P}\right)^{1/2}\left(\frac{\sigma_{\gamma}}{1\,{\rm cm}^2}\right)^{1/2}\\
&\times\left(\frac{3a_B}{\bar{X}}\right)\left(\frac{W_M}{10^{33}\frac{\rm Hz}{e\,{\rm cm}^2}}\right)\left(\frac{M}{10^{-27}\theta_0\,e\,{\rm cm}^2}\right)\,.
\end{aligned}
\end{equation}

We are also interested in the total number of transitions induced in one axion coherence-time $\tau_a$, which may be approximated by the total number of photons absorbed by the target. The latter may be expressed as
\begin{equation}\label{total}
N_{\rm \gamma}=\eta P\tau_al/l_{\rm abs}=\eta P\tau_aln_{\rm mol}\sigma_{\rm abs}\,,
\end{equation}
where $l$ is the length of the sample in the direction of photon propagation, $l_{\rm abs}=1/\left(n_{\rm mol}\sigma_{\rm abs}\right)$ is the photon absorption length, $n_{\rm mol}$ is the molecular density in the target and $\sigma_{\rm abs}$ is the photon absorption cross section. We have included in Eq.\ \eqref{total} the factor $\eta$ to accommodate the possibility of using a cavity to boost the photon flux.

The axion coherence-time may be written as
\begin{equation}\label{no}
\tau_a=10^{6}\times\frac{2\pi}{m_a}=10^{6}\times\frac{2\pi}{\omega}\,,
\end{equation}
where we have imposed the condition that the axion mass is in resonance with the transition frequency.

The cross section $\sigma_{\rm abs}$ is given by [see Eq.\ \eqref{gamma_matrix}]
\begin{equation}\label{cross section}
\sigma_{\rm abs}=\frac{4\pi}{\omega^2}\frac{\Gamma_{\rm abs}}{\Gamma_{\rm tot}}\,\,,\,\,\,\,\,\Gamma_{\rm abs}=\frac{\omega^2}{\left(2\pi\right)^2}\left(\frac{Ze\omega^{3/2}\bar{X}}{Am_p}\right)^2\,,
\end{equation}
where $\Gamma_{\rm tot}$ is the total width of the transition driven by the photons and the axion-induced oscillating nuclear MQM.

Substituting Eqs.\ \eqref{no} and\ \eqref{cross section} into Eq.\ \eqref{total}, we obtain
\begin{equation}
N_{\gamma}=2\times10^6\,\frac{\eta Pln_{\rm mol}}{\Gamma_{\rm tot}}\left(\frac{Ze\omega\bar{X}}{Am_p}\right)^2\,,
\end{equation}
which gives
\begin{equation}
\begin{aligned}
N_{\gamma}&\approx1.7\times10^3\left(\frac{Z}{A}\right)^2\left(\frac{\eta}{10^6}\right)\left(\frac{P}{10^{20}\,{\rm s}^{-1}}\right)\\
&\times\left(\frac{l}{1\,{\rm m}}\right)\left(\frac{10^{-10}\,{\rm eV}}{\Gamma_{\rm tot}}\right)\left(\frac{n_{\rm mol}}{10^{18}\,{\rm cm}^{-3}}\right)\\
&\times\left(\frac{\bar{X}}{3a_B}\right)^2\left(\frac{\omega}{10^{-4}\,{\rm ev}}\right)^2\,.
\end{aligned}
\end{equation}

We note finally that ultimately, the feasibility of the interference scheme discussed here requires at least one axion-induced transition to occur per axion coherence-time. From Eqs.\ \eqref{t_MQM_mol} and\ \eqref{no}, we observe that this, in turn, requires a sample which contains some $10^{30}$ molecules. Achieving such a large number of molecules while maintaining a reasonable sample size is a challenge which may be overcome with future development in experimental techniques.  

\newpage
\bibliography{bib}

\begin{thebibliography}{59}
\expandafter\ifx\csname natexlab\endcsname\relax\def\natexlab#1{#1}\fi
\expandafter\ifx\csname bibnamefont\endcsname\relax
  \def\bibnamefont#1{#1}\fi
\expandafter\ifx\csname bibfnamefont\endcsname\relax
  \def\bibfnamefont#1{#1}\fi
\expandafter\ifx\csname citenamefont\endcsname\relax
  \def\citenamefont#1{#1}\fi
\expandafter\ifx\csname url\endcsname\relax
  \def\url#1{\texttt{#1}}\fi
\expandafter\ifx\csname urlprefix\endcsname\relax\def\urlprefix{URL }\fi
\providecommand{\bibinfo}[2]{#2}
\providecommand{\eprint}[2][]{\url{#2}}

\bibitem[{\citenamefont{Budker et~al.}(2014)\citenamefont{Budker, Graham,
  Ledbetter, Rajendran, and Sushkov}}]{Casper}
\bibinfo{author}{\bibfnamefont{D.}~\bibnamefont{Budker}},
  \bibinfo{author}{\bibfnamefont{P.~W.} \bibnamefont{Graham}},
  \bibinfo{author}{\bibfnamefont{M.}~\bibnamefont{Ledbetter}},
  \bibinfo{author}{\bibfnamefont{S.}~\bibnamefont{Rajendran}},
  \bibnamefont{and} \bibinfo{author}{\bibfnamefont{A.~O.}
  \bibnamefont{Sushkov}}, \bibinfo{journal}{Phys. Rev. X}
  \textbf{\bibinfo{volume}{4}}, \bibinfo{pages}{021030} (\bibinfo{year}{2014}).

\bibitem[{\citenamefont{Stadnik and Flambaum}(2014)}]{Stadnik2014}
\bibinfo{author}{\bibfnamefont{Y.~V.} \bibnamefont{Stadnik}} \bibnamefont{and}
  \bibinfo{author}{\bibfnamefont{V.~V.} \bibnamefont{Flambaum}},
  \bibinfo{journal}{Phys. Rev. D} \textbf{\bibinfo{volume}{89}},
  \bibinfo{pages}{043522} (\bibinfo{year}{2014}).

\bibitem[{\citenamefont{Graham and Rajendran}(2011)}]{Graham}
\bibinfo{author}{\bibfnamefont{P.~W.} \bibnamefont{Graham}} \bibnamefont{and}
  \bibinfo{author}{\bibfnamefont{S.}~\bibnamefont{Rajendran}},
  \bibinfo{journal}{Phys. Rev. D} \textbf{\bibinfo{volume}{84}},
  \bibinfo{pages}{055013} (\bibinfo{year}{2011}).

\bibitem[{\citenamefont{Graham and Rajendran}(2013)}]{GrahamRajendran}
\bibinfo{author}{\bibfnamefont{P.~W.} \bibnamefont{Graham}} \bibnamefont{and}
  \bibinfo{author}{\bibfnamefont{S.}~\bibnamefont{Rajendran}},
  \bibinfo{journal}{Phys. Rev. D} \textbf{\bibinfo{volume}{88}},
  \bibinfo{pages}{035023} (\bibinfo{year}{2013}).

\bibitem[{\citenamefont{Garcon et~al.}(2019)}]{Antoine}
\bibinfo{author}{\bibfnamefont{A.}~\bibnamefont{Garcon}} \bibnamefont{et~al.}
  (\bibinfo{year}{2019}), \eprint{arXiv:1902.04644}.

\bibitem[{\citenamefont{Wu et~al.}(2018)\citenamefont{Wu, Blanchard,
  Jackson~Kimball, Jiang, and Budker}}]{Comagnetometer}
\bibinfo{author}{\bibfnamefont{T.}~\bibnamefont{Wu}},
  \bibinfo{author}{\bibfnamefont{J.~W.} \bibnamefont{Blanchard}},
  \bibinfo{author}{\bibfnamefont{D.~F.} \bibnamefont{Jackson~Kimball}},
  \bibinfo{author}{\bibfnamefont{M.}~\bibnamefont{Jiang}}, \bibnamefont{and}
  \bibinfo{author}{\bibfnamefont{D.}~\bibnamefont{Budker}},
  \bibinfo{journal}{Phys. Rev. Lett.} \textbf{\bibinfo{volume}{121}},
  \bibinfo{pages}{023202} (\bibinfo{year}{2018}).

\bibitem[{\citenamefont{Abel et~al.}(2017)\citenamefont{Abel, Ayres, Ban,
  Bison, Bodek, Bondar, Daum, Fairbairn, Flambaum, Geltenbort et~al.}}]{nEDM}
\bibinfo{author}{\bibfnamefont{C.}~\bibnamefont{Abel}},
  \bibinfo{author}{\bibfnamefont{N.~J.} \bibnamefont{Ayres}},
  \bibinfo{author}{\bibfnamefont{G.}~\bibnamefont{Ban}},
  \bibinfo{author}{\bibfnamefont{G.}~\bibnamefont{Bison}},
  \bibinfo{author}{\bibfnamefont{K.}~\bibnamefont{Bodek}},
  \bibinfo{author}{\bibfnamefont{V.}~\bibnamefont{Bondar}},
  \bibinfo{author}{\bibfnamefont{M.}~\bibnamefont{Daum}},
  \bibinfo{author}{\bibfnamefont{M.}~\bibnamefont{Fairbairn}},
  \bibinfo{author}{\bibfnamefont{V.~V.} \bibnamefont{Flambaum}},
  \bibinfo{author}{\bibfnamefont{P.}~\bibnamefont{Geltenbort}},
  \bibnamefont{et~al.}, \bibinfo{journal}{Phys. Rev. X}
  \textbf{\bibinfo{volume}{7}}, \bibinfo{pages}{041034} (\bibinfo{year}{2017}).

\bibitem[{\citenamefont{Bloch et~al.}(2019)\citenamefont{Bloch, Hochberg,
  Kuflik, and Volansky}}]{VolanskyArxiv}
\bibinfo{author}{\bibfnamefont{I.~M.} \bibnamefont{Bloch}},
  \bibinfo{author}{\bibfnamefont{Y.}~\bibnamefont{Hochberg}},
  \bibinfo{author}{\bibfnamefont{E.}~\bibnamefont{Kuflik}}, \bibnamefont{and}
  \bibinfo{author}{\bibfnamefont{T.}~\bibnamefont{Volansky}}
  (\bibinfo{year}{2019}), \eprint{arXiv:1907.03767}.

\bibitem[{\citenamefont{Smorra et~al.}(2019)\citenamefont{Smorra, Stadnik,
  Blessing, Bohman, Borchert, Devlin, Erlewein, Harrington, Higuchi, Mooser
  et~al.}}]{smorra2019direct}
\bibinfo{author}{\bibfnamefont{C.}~\bibnamefont{Smorra}},
  \bibinfo{author}{\bibfnamefont{Y.}~\bibnamefont{Stadnik}},
  \bibinfo{author}{\bibfnamefont{P.}~\bibnamefont{Blessing}},
  \bibinfo{author}{\bibfnamefont{M.}~\bibnamefont{Bohman}},
  \bibinfo{author}{\bibfnamefont{M.}~\bibnamefont{Borchert}},
  \bibinfo{author}{\bibfnamefont{J.}~\bibnamefont{Devlin}},
  \bibinfo{author}{\bibfnamefont{S.}~\bibnamefont{Erlewein}},
  \bibinfo{author}{\bibfnamefont{J.}~\bibnamefont{Harrington}},
  \bibinfo{author}{\bibfnamefont{T.}~\bibnamefont{Higuchi}},
  \bibinfo{author}{\bibfnamefont{A.}~\bibnamefont{Mooser}},
  \bibnamefont{et~al.}, \bibinfo{journal}{Nature}
  \textbf{\bibinfo{volume}{575}}, \bibinfo{pages}{310} (\bibinfo{year}{2019}).

\bibitem[{\citenamefont{Arvanitaki et~al.}(2018)\citenamefont{Arvanitaki,
  Dimopoulos, and Van~Tilburg}}]{ArvanitakiMolecule}
\bibinfo{author}{\bibfnamefont{A.}~\bibnamefont{Arvanitaki}},
  \bibinfo{author}{\bibfnamefont{S.}~\bibnamefont{Dimopoulos}},
  \bibnamefont{and}
  \bibinfo{author}{\bibfnamefont{K.}~\bibnamefont{Van~Tilburg}},
  \bibinfo{journal}{Phys. Rev. X} \textbf{\bibinfo{volume}{8}},
  \bibinfo{pages}{041001} (\bibinfo{year}{2018}).

\bibitem[{\citenamefont{Crewther et~al.}(1979)\citenamefont{Crewther, Vecchia,
  Veneziano, and Witten}}]{Witten}
\bibinfo{author}{\bibfnamefont{R.}~\bibnamefont{Crewther}},
  \bibinfo{author}{\bibfnamefont{P.~D.} \bibnamefont{Vecchia}},
  \bibinfo{author}{\bibfnamefont{G.}~\bibnamefont{Veneziano}},
  \bibnamefont{and} \bibinfo{author}{\bibfnamefont{E.}~\bibnamefont{Witten}},
  \bibinfo{journal}{Physics Letters B} \textbf{\bibinfo{volume}{88}},
  \bibinfo{pages}{123 } (\bibinfo{year}{1979}).

\bibitem[{\citenamefont{Yamanaka
  et~al.}(2017{\natexlab{a}})\citenamefont{Yamanaka, Sahoo, Yoshinaga, Sato,
  Asahi, and Das}}]{Yamanaka}
\bibinfo{author}{\bibfnamefont{N.}~\bibnamefont{Yamanaka}},
  \bibinfo{author}{\bibfnamefont{B.~K.} \bibnamefont{Sahoo}},
  \bibinfo{author}{\bibfnamefont{N.}~\bibnamefont{Yoshinaga}},
  \bibinfo{author}{\bibfnamefont{T.}~\bibnamefont{Sato}},
  \bibinfo{author}{\bibfnamefont{K.}~\bibnamefont{Asahi}}, \bibnamefont{and}
  \bibinfo{author}{\bibfnamefont{B.~P.} \bibnamefont{Das}},
  \bibinfo{journal}{Eur. Phys. J. A} \textbf{\bibinfo{volume}{53}},
  \bibinfo{pages}{54} (\bibinfo{year}{2017}{\natexlab{a}}).

\bibitem[{\citenamefont{Haxton and Henley}(1983)}]{HH}
\bibinfo{author}{\bibfnamefont{W.~C.} \bibnamefont{Haxton}} \bibnamefont{and}
  \bibinfo{author}{\bibfnamefont{E.~M.} \bibnamefont{Henley}},
  \bibinfo{journal}{Phys. Rev. Lett.} \textbf{\bibinfo{volume}{51}},
  \bibinfo{pages}{1937} (\bibinfo{year}{1983}).

\bibitem[{\citenamefont{Sushkov et~al.}(1984)\citenamefont{Sushkov, Flambaum,
  and Khriplovich}}]{SFK}
\bibinfo{author}{\bibfnamefont{O.}~\bibnamefont{Sushkov}},
  \bibinfo{author}{\bibfnamefont{V.}~\bibnamefont{Flambaum}}, \bibnamefont{and}
  \bibinfo{author}{\bibfnamefont{I.}~\bibnamefont{Khriplovich}},
  \bibinfo{journal}{Sov. Phys. - JETP} \textbf{\bibinfo{volume}{60}},
  \bibinfo{pages}{1521} (\bibinfo{year}{1984}).

\bibitem[{\citenamefont{Flambaum et~al.}(1985)\citenamefont{Flambaum,
  Khriplovich, and Sushkov}}]{FKS1985}
\bibinfo{author}{\bibfnamefont{V.}~\bibnamefont{Flambaum}},
  \bibinfo{author}{\bibfnamefont{I.}~\bibnamefont{Khriplovich}},
  \bibnamefont{and} \bibinfo{author}{\bibfnamefont{O.}~\bibnamefont{Sushkov}},
  \bibinfo{journal}{Phys. Letts. B} \textbf{\bibinfo{volume}{162}},
  \bibinfo{pages}{213 } (\bibinfo{year}{1985}).

\bibitem[{\citenamefont{Flambaum et~al.}(1986)\citenamefont{Flambaum,
  Khriplovich, and Sushkov}}]{FKS1986}
\bibinfo{author}{\bibfnamefont{V.}~\bibnamefont{Flambaum}},
  \bibinfo{author}{\bibfnamefont{I.}~\bibnamefont{Khriplovich}},
  \bibnamefont{and} \bibinfo{author}{\bibfnamefont{O.}~\bibnamefont{Sushkov}},
  \bibinfo{journal}{Nucl. Phys. A} \textbf{\bibinfo{volume}{449}},
  \bibinfo{pages}{750 } (\bibinfo{year}{1986}).

\bibitem[{\citenamefont{Khriplovich and Korkin}(2000)}]{KhriplovichKorkin2000}
\bibinfo{author}{\bibfnamefont{I.}~\bibnamefont{Khriplovich}} \bibnamefont{and}
  \bibinfo{author}{\bibfnamefont{R.}~\bibnamefont{Korkin}},
  \bibinfo{journal}{Nuclear Physics A} \textbf{\bibinfo{volume}{665}},
  \bibinfo{pages}{365 } (\bibinfo{year}{2000}).

\bibitem[{\citenamefont{Yoshinaga et~al.}(2010)\citenamefont{Yoshinaga,
  Higashiyama, and Arai}}]{Yoshinaga2010}
\bibinfo{author}{\bibfnamefont{N.}~\bibnamefont{Yoshinaga}},
  \bibinfo{author}{\bibfnamefont{K.}~\bibnamefont{Higashiyama}},
  \bibnamefont{and} \bibinfo{author}{\bibfnamefont{R.}~\bibnamefont{Arai}},
  \bibinfo{journal}{Prog. Theor. Phys.} \textbf{\bibinfo{volume}{124}},
  \bibinfo{pages}{1115} (\bibinfo{year}{2010}).

\bibitem[{\citenamefont{Mereghetti et~al.}(2011)\citenamefont{Mereghetti,
  de~Vries, Hockings, Maekawa, and van Kolck}}]{MEREGHETTI2011}
\bibinfo{author}{\bibfnamefont{E.}~\bibnamefont{Mereghetti}},
  \bibinfo{author}{\bibfnamefont{J.}~\bibnamefont{de~Vries}},
  \bibinfo{author}{\bibfnamefont{W.}~\bibnamefont{Hockings}},
  \bibinfo{author}{\bibfnamefont{C.}~\bibnamefont{Maekawa}}, \bibnamefont{and}
  \bibinfo{author}{\bibfnamefont{U.}~\bibnamefont{van Kolck}},
  \bibinfo{journal}{Phys. Letts. B} \textbf{\bibinfo{volume}{696}},
  \bibinfo{pages}{97 } (\bibinfo{year}{2011}).

\bibitem[{\citenamefont{de~Vries
  et~al.}(2011{\natexlab{a}})\citenamefont{de~Vries, Mereghetti, Timmermans,
  and van Kolck}}]{deVriesFormFactor2011}
\bibinfo{author}{\bibfnamefont{J.}~\bibnamefont{de~Vries}},
  \bibinfo{author}{\bibfnamefont{E.}~\bibnamefont{Mereghetti}},
  \bibinfo{author}{\bibfnamefont{R.~G.~E.} \bibnamefont{Timmermans}},
  \bibnamefont{and} \bibinfo{author}{\bibfnamefont{U.}~\bibnamefont{van
  Kolck}}, \bibinfo{journal}{Phys. Rev. Lett.} \textbf{\bibinfo{volume}{107}},
  \bibinfo{pages}{091804} (\bibinfo{year}{2011}{\natexlab{a}}).

\bibitem[{\citenamefont{de~Vries
  et~al.}(2011{\natexlab{b}})\citenamefont{de~Vries, Higa, Liu, Mereghetti,
  Stetcu, Timmermans, and van Kolck}}]{deVriesEDM2011}
\bibinfo{author}{\bibfnamefont{J.}~\bibnamefont{de~Vries}},
  \bibinfo{author}{\bibfnamefont{R.}~\bibnamefont{Higa}},
  \bibinfo{author}{\bibfnamefont{C.-P.} \bibnamefont{Liu}},
  \bibinfo{author}{\bibfnamefont{E.}~\bibnamefont{Mereghetti}},
  \bibinfo{author}{\bibfnamefont{I.}~\bibnamefont{Stetcu}},
  \bibinfo{author}{\bibfnamefont{R.~G.~E.} \bibnamefont{Timmermans}},
  \bibnamefont{and} \bibinfo{author}{\bibfnamefont{U.}~\bibnamefont{van
  Kolck}}, \bibinfo{journal}{Phys. Rev. C} \textbf{\bibinfo{volume}{84}},
  \bibinfo{pages}{065501} (\bibinfo{year}{2011}{\natexlab{b}}).

\bibitem[{\citenamefont{Wirzba}(2014)}]{WIRZBA2014}
\bibinfo{author}{\bibfnamefont{A.}~\bibnamefont{Wirzba}},
  \bibinfo{journal}{Nuclear Physics A} \textbf{\bibinfo{volume}{928}},
  \bibinfo{pages}{116 } (\bibinfo{year}{2014}).

\bibitem[{\citenamefont{Dekens et~al.}(2014)\citenamefont{Dekens, de~Vries,
  Bsaisou, Bernreuther, Hanhart, Mei{\ss}ner, Nogga, and Wirzba}}]{Dekens2014}
\bibinfo{author}{\bibfnamefont{W.}~\bibnamefont{Dekens}},
  \bibinfo{author}{\bibfnamefont{J.}~\bibnamefont{de~Vries}},
  \bibinfo{author}{\bibfnamefont{J.}~\bibnamefont{Bsaisou}},
  \bibinfo{author}{\bibfnamefont{W.}~\bibnamefont{Bernreuther}},
  \bibinfo{author}{\bibfnamefont{C.}~\bibnamefont{Hanhart}},
  \bibinfo{author}{\bibfnamefont{U.-G.} \bibnamefont{Mei{\ss}ner}},
  \bibinfo{author}{\bibfnamefont{A.}~\bibnamefont{Nogga}}, \bibnamefont{and}
  \bibinfo{author}{\bibfnamefont{A.}~\bibnamefont{Wirzba}},
  \bibinfo{journal}{J. High Energy Phys.} \textbf{\bibinfo{volume}{2014}},
  \bibinfo{pages}{69} (\bibinfo{year}{2014}).

\bibitem[{\citenamefont{Yoshinaga et~al.}(2014)\citenamefont{Yoshinaga,
  Higashiyama, Arai, and Teruya}}]{YoshinagaXe2014}
\bibinfo{author}{\bibfnamefont{N.}~\bibnamefont{Yoshinaga}},
  \bibinfo{author}{\bibfnamefont{K.}~\bibnamefont{Higashiyama}},
  \bibinfo{author}{\bibfnamefont{R.}~\bibnamefont{Arai}}, \bibnamefont{and}
  \bibinfo{author}{\bibfnamefont{E.}~\bibnamefont{Teruya}},
  \bibinfo{journal}{Phys. Rev. C} \textbf{\bibinfo{volume}{89}},
  \bibinfo{pages}{045501} (\bibinfo{year}{2014}).

\bibitem[{\citenamefont{Bsaisou et~al.}(2015)\citenamefont{Bsaisou, de~Vries,
  Hanhart, Liebig, Mei{\ss}ner, Minossi, Nogga, and Wirzba}}]{Bsaisou2015}
\bibinfo{author}{\bibfnamefont{J.}~\bibnamefont{Bsaisou}},
  \bibinfo{author}{\bibfnamefont{J.}~\bibnamefont{de~Vries}},
  \bibinfo{author}{\bibfnamefont{C.}~\bibnamefont{Hanhart}},
  \bibinfo{author}{\bibfnamefont{S.}~\bibnamefont{Liebig}},
  \bibinfo{author}{\bibfnamefont{U.-G.} \bibnamefont{Mei{\ss}ner}},
  \bibinfo{author}{\bibfnamefont{D.}~\bibnamefont{Minossi}},
  \bibinfo{author}{\bibfnamefont{A.}~\bibnamefont{Nogga}}, \bibnamefont{and}
  \bibinfo{author}{\bibfnamefont{A.}~\bibnamefont{Wirzba}},
  \bibinfo{journal}{J. High Energy Phys.} \textbf{\bibinfo{volume}{2015}},
  \bibinfo{pages}{104} (\bibinfo{year}{2015}).

\bibitem[{\citenamefont{de~Vries et~al.}(2015)\citenamefont{de~Vries,
  Mereghetti, and Walker-Loud}}]{DeVries}
\bibinfo{author}{\bibfnamefont{J.}~\bibnamefont{de~Vries}},
  \bibinfo{author}{\bibfnamefont{E.}~\bibnamefont{Mereghetti}},
  \bibnamefont{and}
  \bibinfo{author}{\bibfnamefont{A.}~\bibnamefont{Walker-Loud}},
  \bibinfo{journal}{Phys. Rev. C} \textbf{\bibinfo{volume}{92}},
  \bibinfo{pages}{045201} (\bibinfo{year}{2015}).

\bibitem[{\citenamefont{Yamanaka and Hiyama}(2015)}]{Yamanaka6Li}
\bibinfo{author}{\bibfnamefont{N.}~\bibnamefont{Yamanaka}} \bibnamefont{and}
  \bibinfo{author}{\bibfnamefont{E.}~\bibnamefont{Hiyama}},
  \bibinfo{journal}{Phys. Rev. C} \textbf{\bibinfo{volume}{91}},
  \bibinfo{pages}{054005} (\bibinfo{year}{2015}).

\bibitem[{\citenamefont{Yamanaka}(2017)}]{YamanakaLightNuclei}
\bibinfo{author}{\bibfnamefont{N.}~\bibnamefont{Yamanaka}},
  \bibinfo{journal}{Int. J. Mod. Phys. A} \textbf{\bibinfo{volume}{26}},
  \bibinfo{pages}{1730002} (\bibinfo{year}{2017}).

\bibitem[{\citenamefont{Yamanaka
  et~al.}(2017{\natexlab{b}})\citenamefont{Yamanaka, Yamada, Hiyama, and
  Funaki}}]{YamanakaPhysRevC2018}
\bibinfo{author}{\bibfnamefont{N.}~\bibnamefont{Yamanaka}},
  \bibinfo{author}{\bibfnamefont{T.}~\bibnamefont{Yamada}},
  \bibinfo{author}{\bibfnamefont{E.}~\bibnamefont{Hiyama}}, \bibnamefont{and}
  \bibinfo{author}{\bibfnamefont{Y.}~\bibnamefont{Funaki}},
  \bibinfo{journal}{Phys. Rev. C} \textbf{\bibinfo{volume}{95}},
  \bibinfo{pages}{065503} (\bibinfo{year}{2017}{\natexlab{b}}).

\bibitem[{\citenamefont{Yamanaka et~al.}(2019)\citenamefont{Yamanaka, Yamada,
  and Funaki}}]{Yamanaka:2019vec}
\bibinfo{author}{\bibfnamefont{N.}~\bibnamefont{Yamanaka}},
  \bibinfo{author}{\bibfnamefont{T.}~\bibnamefont{Yamada}}, \bibnamefont{and}
  \bibinfo{author}{\bibfnamefont{Y.}~\bibnamefont{Funaki}}
  (\bibinfo{year}{2019}), \eprint{arXiv:1907.08091}.

\bibitem[{\citenamefont{Flambaum et~al.}(2014)\citenamefont{Flambaum, DeMille,
  and Kozlov}}]{FDK}
\bibinfo{author}{\bibfnamefont{V.~V.} \bibnamefont{Flambaum}},
  \bibinfo{author}{\bibfnamefont{D.}~\bibnamefont{DeMille}}, \bibnamefont{and}
  \bibinfo{author}{\bibfnamefont{M.~G.} \bibnamefont{Kozlov}},
  \bibinfo{journal}{Phys. Rev. Lett.} \textbf{\bibinfo{volume}{113}},
  \bibinfo{pages}{103003} (\bibinfo{year}{2014}).

\bibitem[{\citenamefont{Chupp et~al.}(2019)\citenamefont{Chupp, Fierlinger,
  Ramsey-Musolf, and Singh}}]{Chupp}
\bibinfo{author}{\bibfnamefont{T.~E.} \bibnamefont{Chupp}},
  \bibinfo{author}{\bibfnamefont{P.}~\bibnamefont{Fierlinger}},
  \bibinfo{author}{\bibfnamefont{M.~J.} \bibnamefont{Ramsey-Musolf}},
  \bibnamefont{and} \bibinfo{author}{\bibfnamefont{J.~T.} \bibnamefont{Singh}},
  \bibinfo{journal}{Rev. Mod. Phys.} \textbf{\bibinfo{volume}{91}},
  \bibinfo{pages}{015001} (\bibinfo{year}{2019}).

\bibitem[{\citenamefont{Flambaum and Ginges}(2002)}]{FlambaumGinges2002}
\bibinfo{author}{\bibfnamefont{V.~V.} \bibnamefont{Flambaum}} \bibnamefont{and}
  \bibinfo{author}{\bibfnamefont{J.~S.~M.} \bibnamefont{Ginges}},
  \bibinfo{journal}{Phys. Rev. A} \textbf{\bibinfo{volume}{65}},
  \bibinfo{pages}{032113} (\bibinfo{year}{2002}).

\bibitem[{\citenamefont{Jesus and Engel}(2005)}]{JesusHg2005}
\bibinfo{author}{\bibfnamefont{J.~H.~d.} \bibnamefont{Jesus}} \bibnamefont{and}
  \bibinfo{author}{\bibfnamefont{J.}~\bibnamefont{Engel}},
  \bibinfo{journal}{Phys. Rev. C} \textbf{\bibinfo{volume}{72}},
  \bibinfo{pages}{045503} (\bibinfo{year}{2005}).

\bibitem[{\citenamefont{Cho et~al.}(1991)\citenamefont{Cho, Sangster, and
  Hinds}}]{ChoTl1991}
\bibinfo{author}{\bibfnamefont{D.}~\bibnamefont{Cho}},
  \bibinfo{author}{\bibfnamefont{K.}~\bibnamefont{Sangster}}, \bibnamefont{and}
  \bibinfo{author}{\bibfnamefont{E.~A.} \bibnamefont{Hinds}},
  \bibinfo{journal}{Phys. Rev. A} \textbf{\bibinfo{volume}{44}},
  \bibinfo{pages}{2783} (\bibinfo{year}{1991}).

\bibitem[{\citenamefont{Griffith et~al.}(2009)\citenamefont{Griffith, Swallows,
  Loftus, Romalis, Heckel, and Fortson}}]{GriffithHg2011}
\bibinfo{author}{\bibfnamefont{W.~C.} \bibnamefont{Griffith}},
  \bibinfo{author}{\bibfnamefont{M.~D.} \bibnamefont{Swallows}},
  \bibinfo{author}{\bibfnamefont{T.~H.} \bibnamefont{Loftus}},
  \bibinfo{author}{\bibfnamefont{M.~V.} \bibnamefont{Romalis}},
  \bibinfo{author}{\bibfnamefont{B.~R.} \bibnamefont{Heckel}},
  \bibnamefont{and} \bibinfo{author}{\bibfnamefont{E.~N.}
  \bibnamefont{Fortson}}, \bibinfo{journal}{Phys. Rev. Lett.}
  \textbf{\bibinfo{volume}{102}}, \bibinfo{pages}{101601}
  (\bibinfo{year}{2009}).

\bibitem[{\citenamefont{Graner et~al.}(2016)\citenamefont{Graner, Chen,
  Lindahl, and Heckel}}]{GranerHg2017}
\bibinfo{author}{\bibfnamefont{B.}~\bibnamefont{Graner}},
  \bibinfo{author}{\bibfnamefont{Y.}~\bibnamefont{Chen}},
  \bibinfo{author}{\bibfnamefont{E.~G.} \bibnamefont{Lindahl}},
  \bibnamefont{and} \bibinfo{author}{\bibfnamefont{B.~R.}
  \bibnamefont{Heckel}}, \bibinfo{journal}{Phys. Rev. Lett.}
  \textbf{\bibinfo{volume}{116}}, \bibinfo{pages}{161601}
  (\bibinfo{year}{2016}).

\bibitem[{\citenamefont{Auerbach et~al.}(1996)\citenamefont{Auerbach, Flambaum,
  and Spevak}}]{Auberbach1996}
\bibinfo{author}{\bibfnamefont{N.}~\bibnamefont{Auerbach}},
  \bibinfo{author}{\bibfnamefont{V.~V.} \bibnamefont{Flambaum}},
  \bibnamefont{and} \bibinfo{author}{\bibfnamefont{V.}~\bibnamefont{Spevak}},
  \bibinfo{journal}{Phys. Rev. Lett.} \textbf{\bibinfo{volume}{76}},
  \bibinfo{pages}{4316} (\bibinfo{year}{1996}).

\bibitem[{\citenamefont{Spevak et~al.}(1997)\citenamefont{Spevak, Auerbach, and
  Flambaum}}]{Spevak1997}
\bibinfo{author}{\bibfnamefont{V.}~\bibnamefont{Spevak}},
  \bibinfo{author}{\bibfnamefont{N.}~\bibnamefont{Auerbach}}, \bibnamefont{and}
  \bibinfo{author}{\bibfnamefont{V.~V.} \bibnamefont{Flambaum}},
  \bibinfo{journal}{Phys. Rev. C} \textbf{\bibinfo{volume}{56}},
  \bibinfo{pages}{1357} (\bibinfo{year}{1997}).

\bibitem[{\citenamefont{Engel et~al.}(2003)\citenamefont{Engel, Bender,
  Dobaczewski, Jesus, and Olbratowski}}]{EngelRa2003}
\bibinfo{author}{\bibfnamefont{J.}~\bibnamefont{Engel}},
  \bibinfo{author}{\bibfnamefont{M.}~\bibnamefont{Bender}},
  \bibinfo{author}{\bibfnamefont{J.}~\bibnamefont{Dobaczewski}},
  \bibinfo{author}{\bibfnamefont{J.~H.~d.} \bibnamefont{Jesus}},
  \bibnamefont{and}
  \bibinfo{author}{\bibfnamefont{P.}~\bibnamefont{Olbratowski}},
  \bibinfo{journal}{Phys. Rev. C} \textbf{\bibinfo{volume}{68}},
  \bibinfo{pages}{025501} (\bibinfo{year}{2003}).

\bibitem[{\citenamefont{Dobaczewski et~al.}(2018)\citenamefont{Dobaczewski,
  Engel, Kortelainen, and Becker}}]{Dobaczewski2018}
\bibinfo{author}{\bibfnamefont{J.}~\bibnamefont{Dobaczewski}},
  \bibinfo{author}{\bibfnamefont{J.}~\bibnamefont{Engel}},
  \bibinfo{author}{\bibfnamefont{M.}~\bibnamefont{Kortelainen}},
  \bibnamefont{and} \bibinfo{author}{\bibfnamefont{P.}~\bibnamefont{Becker}},
  \bibinfo{journal}{Phys. Rev. Lett.} \textbf{\bibinfo{volume}{121}},
  \bibinfo{pages}{232501} (\bibinfo{year}{2018}).

\bibitem[{\citenamefont{Flambaum and Feldmeier}(2019)}]{Flambaum2019tym}
\bibinfo{author}{\bibfnamefont{V.~V.} \bibnamefont{Flambaum}} \bibnamefont{and}
  \bibinfo{author}{\bibfnamefont{H.}~\bibnamefont{Feldmeier}}
  (\bibinfo{year}{2019}), \eprint{arXiv:1907.07438}.

\bibitem[{\citenamefont{Flambaum and Dzuba}(2019)}]{flambaum2019electric}
\bibinfo{author}{\bibfnamefont{V.~V.} \bibnamefont{Flambaum}} \bibnamefont{and}
  \bibinfo{author}{\bibfnamefont{V.~A.} \bibnamefont{Dzuba}}
  (\bibinfo{year}{2019}), \eprint{arXiv:1912.03598}.

\bibitem[{\citenamefont{Skripnikov et~al.}(2017)\citenamefont{Skripnikov,
  Titov, and Flambaum}}]{Skripnikov2017}
\bibinfo{author}{\bibfnamefont{L.~V.} \bibnamefont{Skripnikov}},
  \bibinfo{author}{\bibfnamefont{A.~V.} \bibnamefont{Titov}}, \bibnamefont{and}
  \bibinfo{author}{\bibfnamefont{V.~V.} \bibnamefont{Flambaum}},
  \bibinfo{journal}{Phys. Rev. A} \textbf{\bibinfo{volume}{95}},
  \bibinfo{pages}{022512} (\bibinfo{year}{2017}).

\bibitem[{\citenamefont{Flambaum}(1994)}]{FlambaumMQM1994}
\bibinfo{author}{\bibfnamefont{V.}~\bibnamefont{Flambaum}},
  \bibinfo{journal}{Physics Letters B} \textbf{\bibinfo{volume}{320}},
  \bibinfo{pages}{211 } (\bibinfo{year}{1994}).

\bibitem[{\citenamefont{Lackenby and Flambaum}(2018)}]{Lackenby2018}
\bibinfo{author}{\bibfnamefont{B.~G.~C.} \bibnamefont{Lackenby}}
  \bibnamefont{and} \bibinfo{author}{\bibfnamefont{V.~V.}
  \bibnamefont{Flambaum}}, \bibinfo{journal}{Phys. Rev. D}
  \textbf{\bibinfo{volume}{98}}, \bibinfo{pages}{115019}
  (\bibinfo{year}{2018}).

\bibitem[{\citenamefont{Centers et~al.}(2019)}]{Centers:2019dyn}
\bibinfo{author}{\bibfnamefont{G.~P.} \bibnamefont{Centers}}
  \bibnamefont{et~al.} (\bibinfo{year}{2019}), \eprint{arXiv: 1905.13650}.

\bibitem[{\citenamefont{Landau and Lifshitz}(1965)}]{Landau}
\bibinfo{author}{\bibfnamefont{L.}~\bibnamefont{Landau}} \bibnamefont{and}
  \bibinfo{author}{\bibfnamefont{E.}~\bibnamefont{Lifshitz}},
  \emph{\bibinfo{title}{{Quantum Mechanics}}} (\bibinfo{publisher}{Pergamon
  press}, \bibinfo{address}{Oxford}, \bibinfo{year}{1965}).

\bibitem[{\citenamefont{Bethe and Salpeter}(1957)}]{Bethe}
\bibinfo{author}{\bibfnamefont{H.}~\bibnamefont{Bethe}} \bibnamefont{and}
  \bibinfo{author}{\bibfnamefont{E.}~\bibnamefont{Salpeter}},
  \emph{\bibinfo{title}{Quantum Mechanics of One- and Two-Electron Atoms}}
  (\bibinfo{publisher}{Springer}, \bibinfo{year}{1957}).

\bibitem[{\citenamefont{Dzuba et~al.}(2000)\citenamefont{Dzuba, Flambaum, and
  Ginges}}]{DzubaFlambaumGinges2000}
\bibinfo{author}{\bibfnamefont{V.~A.} \bibnamefont{Dzuba}},
  \bibinfo{author}{\bibfnamefont{V.~V.} \bibnamefont{Flambaum}},
  \bibnamefont{and} \bibinfo{author}{\bibfnamefont{J.~S.~M.}
  \bibnamefont{Ginges}}, \bibinfo{journal}{Phys. Rev. A}
  \textbf{\bibinfo{volume}{61}}, \bibinfo{pages}{062509}
  (\bibinfo{year}{2000}).

\bibitem[{\citenamefont{Klaer and Moore}(2017)}]{AxionMass2017}
\bibinfo{author}{\bibfnamefont{V.~B.} \bibnamefont{Klaer}} \bibnamefont{and}
  \bibinfo{author}{\bibfnamefont{G.~D.} \bibnamefont{Moore}},
  \bibinfo{journal}{J. Cosmol. Astropart. Phys.}
  \textbf{\bibinfo{volume}{2017}}, \bibinfo{pages}{049} (\bibinfo{year}{2017}).

\bibitem[{\citenamefont{Buschmann et~al.}(2019)\citenamefont{Buschmann, Foster,
  and Safdi}}]{Buschmann:2019icd}
\bibinfo{author}{\bibfnamefont{M.}~\bibnamefont{Buschmann}},
  \bibinfo{author}{\bibfnamefont{J.~W.} \bibnamefont{Foster}},
  \bibnamefont{and} \bibinfo{author}{\bibfnamefont{B.~R.} \bibnamefont{Safdi}}
  (\bibinfo{year}{2019}), \eprint{arXiv: 1906.00967}.

\bibitem[{\citenamefont{Gorghetto et~al.}(2018)\citenamefont{Gorghetto, Hardy,
  and Villadoro}}]{Gorghetto_2018}
\bibinfo{author}{\bibfnamefont{M.}~\bibnamefont{Gorghetto}},
  \bibinfo{author}{\bibfnamefont{E.}~\bibnamefont{Hardy}}, \bibnamefont{and}
  \bibinfo{author}{\bibfnamefont{G.}~\bibnamefont{Villadoro}},
  \bibinfo{journal}{J. High Energy Phys.} \textbf{\bibinfo{volume}{2018}},
  \bibinfo{pages}{151} (\bibinfo{year}{2018}).

\bibitem[{\citenamefont{Tan et~al.}(2019)\citenamefont{Tan, Flambaum, and
  Samsonov}}]{TranFlambaum2019}
\bibinfo{author}{\bibfnamefont{H.~B.~T.} \bibnamefont{Tan}},
  \bibinfo{author}{\bibfnamefont{V.~V.} \bibnamefont{Flambaum}},
  \bibnamefont{and} \bibinfo{author}{\bibfnamefont{I.~B.}
  \bibnamefont{Samsonov}}, \bibinfo{journal}{Phys. Rev. A}
  \textbf{\bibinfo{volume}{99}}, \bibinfo{pages}{013430}
  (\bibinfo{year}{2019}).

\bibitem[{\citenamefont{Flambaum}(2019)}]{Flambaum229Th2019}
\bibinfo{author}{\bibfnamefont{V.~V.} \bibnamefont{Flambaum}},
  \bibinfo{journal}{Phys. Rev. C} \textbf{\bibinfo{volume}{99}},
  \bibinfo{pages}{035501} (\bibinfo{year}{2019}).

\bibitem[{\citenamefont{Kozlov et~al.}(1987)\citenamefont{Kozlov, Fomichev,
  Dmitriev, Labzovsky, and Titov}}]{Kozlov_1987}
\bibinfo{author}{\bibfnamefont{M.~G.} \bibnamefont{Kozlov}},
  \bibinfo{author}{\bibfnamefont{V.~I.} \bibnamefont{Fomichev}},
  \bibinfo{author}{\bibfnamefont{Y.~Y.} \bibnamefont{Dmitriev}},
  \bibinfo{author}{\bibfnamefont{L.~N.} \bibnamefont{Labzovsky}},
  \bibnamefont{and} \bibinfo{author}{\bibfnamefont{A.~V.} \bibnamefont{Titov}},
  \bibinfo{journal}{J. Phys. B} \textbf{\bibinfo{volume}{20}},
  \bibinfo{pages}{4939} (\bibinfo{year}{1987}).

\bibitem[{\citenamefont{Dmitriev et~al.}(1992)\citenamefont{Dmitriev, Khait,
  Kozlov, Labzovsky, Mitrushenkov, Shtoff, and Titov}}]{DMITRIEV1992}
\bibinfo{author}{\bibfnamefont{Y.}~\bibnamefont{Dmitriev}},
  \bibinfo{author}{\bibfnamefont{Y.}~\bibnamefont{Khait}},
  \bibinfo{author}{\bibfnamefont{M.}~\bibnamefont{Kozlov}},
  \bibinfo{author}{\bibfnamefont{L.}~\bibnamefont{Labzovsky}},
  \bibinfo{author}{\bibfnamefont{A.}~\bibnamefont{Mitrushenkov}},
  \bibinfo{author}{\bibfnamefont{A.}~\bibnamefont{Shtoff}}, \bibnamefont{and}
  \bibinfo{author}{\bibfnamefont{A.}~\bibnamefont{Titov}},
  \bibinfo{journal}{Phys. Letts. A} \textbf{\bibinfo{volume}{167}},
  \bibinfo{pages}{280 } (\bibinfo{year}{1992}), ISSN \bibinfo{issn}{0375-9601}.

\bibitem[{\citenamefont{Kozlov and Labzowsky}(1995)}]{Kozlov_1995}
\bibinfo{author}{\bibfnamefont{M.~G.} \bibnamefont{Kozlov}} \bibnamefont{and}
  \bibinfo{author}{\bibfnamefont{L.~N.} \bibnamefont{Labzowsky}},
  \bibinfo{journal}{J. Phys. B} \textbf{\bibinfo{volume}{28}},
  \bibinfo{pages}{1933} (\bibinfo{year}{1995}).

\bibitem[{\citenamefont{Rochester et~al.}(2016)\citenamefont{Rochester,
  Szyma\ifmmode~\acute{n}\else \'{n}\fi{}ski, Raizen, Pustelny, Auzinsh, and
  Budker}}]{DimaCooling}
\bibinfo{author}{\bibfnamefont{S.~M.} \bibnamefont{Rochester}},
  \bibinfo{author}{\bibfnamefont{K.}~\bibnamefont{Szyma\ifmmode~\acute{n}\else
  \'{n}\fi{}ski}}, \bibinfo{author}{\bibfnamefont{M.}~\bibnamefont{Raizen}},
  \bibinfo{author}{\bibfnamefont{S.}~\bibnamefont{Pustelny}},
  \bibinfo{author}{\bibfnamefont{M.}~\bibnamefont{Auzinsh}}, \bibnamefont{and}
  \bibinfo{author}{\bibfnamefont{D.}~\bibnamefont{Budker}},
  \bibinfo{journal}{Phys. Rev. A} \textbf{\bibinfo{volume}{94}},
  \bibinfo{pages}{043416} (\bibinfo{year}{2016}).

\end{thebibliography}

\end{document}